\documentclass[a4paper,usenatbib]{mn2e}

\usepackage{hyperref}
\usepackage{amsfonts}
\usepackage{amssymb}
\usepackage{amsmath}
\usepackage{graphicx}
\usepackage[totalwidth=506pt,totalheight=680pt]{geometry}

\newcommand{\adsurl}[1]{\href{#1}{ADS}}
\providecommand{\url}[1]{\href{#1}{#1}}

\newcommand{\avg}[1]{\ensuremath{\langle \,#1\, \rangle}}

\newcommand{\eqn}[1]{equation~\eqref{#1}}

\newcommand{\dd}{\mathrm{d}}
\newcommand{\be}{\begin{equation}}
\newcommand{\ee}{\end{equation}}
\newcommand{\D}{\mathcal{D}}
\newcommand{\G}{\mathrm{G}}

\newcommand{\PS}{\mathrm{CC}}
\newcommand{\up}{\mathrm{up}}
\renewcommand{\L}{\mathrm{L}}
\newcommand{\DL}{\Delta_\mathrm{L}}
\newcommand{\BL}{B_\mathrm{L}}
\newcommand{\pd}{\partial}

\title[Excursion sets in non-Gaussian random fields]
      {The excursion set approach in non-Gaussian random fields}

\author[M.~Musso, R.~K.~Sheth]{%
Marcello Musso$^{1}$\thanks{E-mail: marcello.musso@uclouvain.be} 
\& Ravi K.~Sheth$^{2,3}$\thanks{E-mail: sheth@ictp.it}\\
 $^1$ CP3-IRMP, Universit\'e Catholique de Louvain, 
      2 Chemin du Cyclotron, 1348 Louvain-la-Neuve, Belgium \\
 $^2$ The Abdus Salam International Center for Theoretical Physics,
      Strada Costiera, 11, Trieste 34151, Italy\\
 $^3$ Center for Particle Cosmology, University of Pennsylvania, 
      209 S. 33rd St., Philadelphia, PA 19104, USA}

\begin{document}

\pagerange{\pageref{firstpage}--\pageref{lastpage}}

\maketitle 

\label{firstpage}

\begin{abstract}

Insight into a number of interesting questions in cosmology can be obtained by studying the first crossing distributions of physically motivated barriers by random walks with correlated steps:  higher mass objects are associated with walks that cross the barrier in fewer steps.  We write the first crossing distribution as a formal series, ordered by the number of times a walk upcrosses the barrier.  Since the fraction of walks with many upcrossings is negligible if the walk has not taken many steps, the leading order term in this series is the most relevant for understanding the massive objects of most interest in cosmology.  For walks associated with Gaussian random fields, this first term only requires knowledge of the bivariate distribution of the walk height and slope, and provides an excellent approximation to the first crossing distribution for all barriers and smoothing filters of current interest.  We show that this simplicity survives when extending the approach to the case of non-Gaussian random fields.  For non-Gaussian fields which are obtained by deterministic transformations of a Gaussian, the first crossing distribution is simply related to that for Gaussian walks crossing a suitably rescaled barrier.  Our analysis shows that this is a useful way to think of the generic case as well.
Although our study is motivated by the possibility that the primordial fluctuation field was non-Gaussian, our results are general.  In particular, they do not assume the non-Gaussianity is small, so they may be viewed as the solution to an excursion set analysis of the late-time, nonlinear fluctuation field rather than the initial one.  They are also useful for models in which the barrier height is determined by quantities other than the initial density, since most other physically motivated variables (such as the shear) are usually stochastic and non-Gaussian.  We use the Lognormal transformation to illustrate some of our arguments.  
\end{abstract}


%%%%%%%%%%%%%%%%%%%%%%%%%%%%%%%%%%%%%%%%%%%%%%%%%%%%%%%%%%%%%
%%%%%%%%%%%%%%%%%%%%%%%%%%%%%%%%%%%%%%%%%%%%%%%%%%%%%%%%%%%%%

\section{Introduction}

%%%%%%%%%%%%%%%%%%%%%%%%%%%%%%%%%%%%%%%%%%%%%%%%%%%%%%%%%%%%%

The statistical distribution of gravitationally bound objects in the Universe is a powerful tool for constraining the amount of primordial non-Gaussianity, thus helping shed some light on the physics of the very early times. The dependence on mass of the abundance and spatial correlations of collapsed objects are useful and complementary tools for probing non-Gaussianity on different scales, in particular, scales that are smaller than those accessible with CMB observations.

The excursion set approach \citep{bcek91} provides an analytical framework for linking the statistics of haloes to fluctuations in the primordial density field. In this approach, one studies the overdensity field $\delta$ smoothed on the scale $R$,
\begin{equation}
  \delta(\mathbf{x},R)=\int \dd \mathbf{y}\,
  W_R(\mathbf{x}-\mathbf{y}) \delta (\mathbf{y})\,,
\end{equation}
where $\mathbf{x}$ and $\mathbf{y}$ are spatial coordinates and $W_R$ is a filter that goes to zero for $|\mathbf{x}-\mathbf{y}|\gg R$. At a given (randomly chosen) position in space the evolution of $\delta_R$ as a function of (the inverse of) $R$ resembles a random trajectory. Repeating this for every position in space gives an ensemble of trajectories, each starting from zero (homogeneity demands $\delta=0$ for infinitely large smoothing scales). For each trajectory, one looks for the largest $R$ (if any) for which the value of the smoothed density field lies above some threshold value (which may itself depend on $R$). An object of mass $M \sim  R^3$ is then associated with that trajectory.  

If $\dd n/\dd M$ denotes the comoving number density of haloes of mass $M$, then the mass fraction in such halos is $(M/\bar\rho)\,\dd n/\dd M$, where $\bar\rho$ is the comoving background density.  The excursion set approach assumes that this mass fraction equals the fraction of walks which cross the threshold (the ``barrier'') for the first time when the smoothing scale is $R$:
\begin{equation}
  f(R)\,\dd R = (M/\bar\rho)\,(\dd n/\dd M)\, \dd M.
\label{ansatz}
\end{equation}
Although recent work has focussed on the shortcomings of this ansatz \citep{bm96,ps12}, the first crossing distribution is nevertheless expected to provide substantial insight into the dependence of $\dd n/\dd M$ on cosmological parameters.  In any case, the question of how the first crossing distribution depends on the nature of the underlying fluctuation field is interesting in its own right.

A crucial part of the problem is to avoid double counting trajectories, i.e., to discard at lower scales all trajectories that have already crossed at larger scales (since they are already associated with an object of larger mass -- given by the largest scale on which the trajectory crossed the barrier). This is rather straightforward to implement numerically, but hard to deal with analytically. Indeed, exact solutions are known only for the unrealistic case of walks with uncorrelated steps (for Gaussian fields, this corresponds to a smoothing filter that is a sharp step function in Fourier space) and only for a few specific barriers.  Considerable effort has been devoted to finding satisfactory analytical approximations, or fitting formulae, for the generic case in which steps are correlated.

The problem is potentially even harder for non-Gaussian initial conditions, since different Fourier modes of the field become coupled, and this introduces additional correlations between the steps, whatever the smoothing filter. Moreover, the most sizeable non-Gaussian deviations are likely to be in the (massive object) tail of the distribution. In this regime, perturbative expansions around the Gaussian result are likely to blow up, so they must be handled with care \citep{damico11}.

In this paper we provide a simple analytic approximation scheme that works for a broad variety of barriers and filters, and can be implemented up to an arbitrary precision level for any (Gaussian or non-Gaussian) distribution of the underlying matter density field. The general formalism is presented in Section~\ref{sec:formalism}, and explicit calculations are carried out in Section~\ref{sec:main}, where we summarize our previous work on Gaussian fields and show how to extend it to non-Gaussian fields.  Section~\ref{sec:other} shows how to use our results when stochastic (non-Gaussian) variables other than the initial overdensity determine halo formation, and as the basis of an excursion set study based on the late-time, nonlinear (rather than initial) fluctuation field.  A final section summarizes our results.  Appendix~\ref{sec:corrections} discusses how to go beyond the simplest approximation we present in the main text, and our use of the Edgeworth and related-expansions for approximating non-Gaussian distributions is summarized in Appendix~\ref{sec:pdf}.

%%%%%%%%%%%%%%%%%%%%%%%%%%%%%%%%%%%%%%%%%%%%%%%%%%%%%%%%%%%%

\section{First crossing distribution with correlated steps}
\label{sec:formalism}

%%%%%%%%%%%%%%%%%%%%%%%%%%%%%%%%%%%%%%%%%%%%%%%%%%%%%%%%%%%%

In hierarchical models, the variance $s\equiv\avg{\delta^2(R)}$ of the density field $\delta$ when smoothed on scale $R$ vanishes by definition for $R=\infty$, and it grows monotonically for smaller $R$ (note that $\avg{\!\delta\!}\equiv 0$ for any $R$), according to
\begin{equation}
  s(R) = \int \frac{\dd k}{k}\, \frac{k^3P(k)}{2\pi^2} \, W^2(kR)\,,  
\end{equation}
where $P(k)$ is the power spectrum of $\delta$.
Therefore, $R$ and $s$ can be used interchangeably, and it is customary and convenient to study the walks as a function of $s$ rather than $R$, as this has the advantage of hiding the dependence on the power spectrum and the smoothing filter.  These only appear when the actual relation between $s$ and $R$ is needed.

What we are after is the first crossing rate, i.e.~the probability that a walk $\delta$ crosses for the first time the barrier $b(s)$ at some scale $s$. In other words, we want to compute the probability that $\delta(s) > b(s)$ at $s$ but $\delta(s_1) < b(s_1)$ for all $s_1<s$, knowing the probability distribution $p(\delta;s)$ of the walk values at any $s$.  In general, requiring $\delta(s) > b(s)$ is straightforward, whereas the additional constraint on the walk heights for all $s_1<s$ is difficult to treat analytically.

\subsection{Height alone}

In one of the earliest works on this subject, \citet{ps74} simply ignored this constraint, and estimated $f(s)$ as
\begin{equation}
\label{fPS}
  f_\mathrm{CC}(s) = -
  \frac{\dd}{\dd s} \int^{b(s)}_{-\infty} {\rm d}\delta\,p(\delta;s) \, .
\end{equation}
(The reason for the subscript CC will become clear shortly.)  Strictly speaking, Press \& Schechter ended up multiplying the right hand side of this expression by a factor of 2, and they only studied the special case in which $b$ is independent of $s$.  In their case, $b\equiv\delta_c=1.686$ was the threshold inferred from spherical collapse.
The extension to barriers which decrease monotonically as $s$ increases is trivial; otherwise %if the barrier increases sufficiently rapidly with $s$, then 
one must be a little more careful, as we discuss below.  That this formulation does not impose any constraint on the walk values at larger scales (smaller $s$) is a point which was highlighted by \cite{bcek91}.  In fact, it does not even distinguish between trajectories crossing the threshold upwards or downwards, a point to which we return shortly.

Recently \citet*{pls12} noted that there is an interesting and instructive limit in which \eqn{fPS} is exact: the set of smooth deterministic curves having $\delta\propto\sqrt{s}$.  Each of these curves represents what they called a completely correlated walk, that is a monotonic function of $s$ whose amplitude is set by a single number, the constant of proportionality, which one may take to be the height on scale $s=1$.  If the distribution of this constant is specified on one scale (say $s=1$) then the distribution of $\delta$ on another scale, $p(\delta;s)$, is simply related to $p(\delta;1)$, and, for this family of curves, \eqn{fPS} is exact:  hence the subscript CC.  This limit is interesting because, regardless of the filter and the matter power spectrum, all correlated walks tend to deterministic trajectories with $\delta\propto\sqrt{s}$ as $s\to0$.  Thus, in this (large mass) limit, \eqn{fPS} is exact, explaining the numerical results of \cite{bcek91}.

At larger values of $s$, the completely correlated limit is no longer so accurate.  However, although small fluctuations around these trajectories appear, most walks still remain monotonic functions of $s$.  Therefore the contribution to $p(\delta;s)$ from walks criss-crossing the barrier multiple times is still negligible, so the constraint $\delta(s_1) < b(s_1)$ for $s_1<s$ is automatically satisfied.

\subsection{Upcrossing requires both height and slope}

As $s$ gets larger, one must account for larger and larger fluctuations away from the deterministic trajectories.  The most efficient way of doing so, at fixed height $\delta$ on scale $s$, is to consider fluctuations in the slope $v\equiv \dd \delta/\dd s$ ($v$ for velocity) \citep{ms12}.  (For the ensemble of deterministic walks, the distribution of the slope $v$ at fixed height $\delta$ is a delta-function centered on $\delta/2s$.)  Since one only wants to count walks that are crossing the barrier upwards, to the condition that $\delta = b(s)$ one should add the requirement that $v \ge \dd b/\dd s$ (for a barrier of constant height, this is just $v\ge 0$).  
%can be obtained by adding a constraint on the increment of $\delta$ over the infinitesimal interval $\Delta s$ (the ``velocity'' of the walk). Only walks with $\delta=b(s)$ and an increment rate larger than $\mathrm{d}b/\mathrm{d}s$ (the velocity of the barrier) should be included.  
%All other walks are crossing downwards at $s$ and must be discarded, since they were necessarily above threshold between some $s_1<s$ and $s$. Clearly, this criterion fails to discard those walks that stayed below threshold between $s_1$ and $s$, but had crossed it at $s_2<s_1$, i.e.~walks with more than one upcrossing.
%for which $\delta(s_1)<b(s_1)$ but $\delta(s_2)>b(s_2)$. 

Thus, if earlier upcrossings can be neglected, $f(s)$ can be computed from the joint probability $p(\delta,v;s)$ that a walk reaches $\delta$ at scale $s$ with velocity $v$, as
\begin{equation}
  f(s) \simeq f_\up(s) \equiv
     \int_{b'(s)}^{+\infty} \!\!\mathrm{d}v\, [v-b'(s)]\, p (b(s),v;s)\,.
\label{fmeanv}
\end{equation}
Clearly, this formulation fails to discard those walks that were above threshold at some $s_1<s$, but with $\delta(s_2)<b(s_2)$ for $s_1<s_2<s$, i.e.~walks with more than one upcrossing.  However, at small $s$, the fraction of such walks is tiny, since the correlations between the steps make sharp turns very unlikely. 

Using Dirac's and Heaviside's distributions $\delta_{\mathrm{D}}$ and $\vartheta$, and since $p(b,v)=\avg{\delta_{\mathrm{D}}(\delta-b)\delta_{\mathrm{D}}(\delta'-v)}$, this approximation can also be written as
\begin{equation}
\label{fstheta}
  f(s) \simeq f_\up(s) = 
  \bigg\langle{\bigg[\frac{\mathrm{d}}{\mathrm{d}s} \vartheta(\delta-b)\bigg]
  \vartheta(\delta'-b')} \bigg\rangle\,,
\end{equation}
which makes it clear that the condition to recover $f_{\PS}$ is that $\delta'>b'$ for most realisations. This is exactly the case for correlated steps in the large mass regime, since $\delta=b$ implies that typically $\delta'\sim b/s $, and $b'\ll b/s$ for small $s$ (as long as the barrier is not receeding too fast from its initial value).

In terms of the conditional probability $p(v|b(s))=p(b(s),v;s)/p(b(s);s)$, and omitting for ease of notation all explicit $s$ dependences, the upcrossing rate also reads
\begin{equation}
  f_\up(s) = p(b) %  \avg{\!(v-b')\vartheta(v-b')|b},
  \!\int_{b'}^{+\infty} \!\!\mathrm{d}v \, (v-b') \, p (v|b) \,.
\label{fmnv}
\end{equation}
This allows a very intuitive explanation in terms of particles in a box: $p(b)$ plays the r\^ole of a number density at $b$ (the number of particles in the one-dimensional volume element $\mathrm{d}\delta$), while the integral is the mean of $\delta'-b'$ over all velocities larger than the barrier's increment given that $\delta=b$, that is the average escape velocity at $b$. The product of the two evaluated at the boundary is by definition the escape rate from the box.
This makes it also easy to see the connection to deterministic walks for which $p(v|b)\to \delta_{\rm D}(b/2s)$, and thus $f_\up(s)\to p(b)\,(b/2s - b') = -p(b/\sqrt{s})\, \dd (b/\sqrt{s})/\dd s$, which is indeed $f_\mathrm{CC}(s)$.  Of course, at larger $s$, when $p(v|\delta)$ is broader, $f_\up(s)$ is a substantially more accurate approximation for $f(s)$.

\subsection{Accounting for multiple upcrossings}

\label{multiple}

The approximation of \eqn{fmeanv} accounts for all walks that cross the barrier upwards at $s$, including those that crossed the barrier previously, and thus it overestimates $f(s)$. The error is expected to increase as $s$ gets large, when such walks become increasingly common.  Removing all the walks that crossed at $s_1<s$, i.e. walks with $\delta(s_1)=b(s_1)$ and $v(s_1)>b'(s_1)$, and then integrating over $s_1$, would account correctly for the trajectories with just one crossing before the last one. So, if we stopped here, and assuming for simplicity a constant barrier, then we would get 
\begin{align}
\label{eq:corrections}
  f(s) &= \int_0^{+\infty} \!\!\mathrm{d}v \, v \, p (b,v) \notag \\
       &  - \int_0^s\!\dd s_1\int_0^{\infty} \!\!\dd v_1\, v_1
            \int_0^{\infty} \!\!\dd v\, v\,p (b_1,v_1,b,v)+\dots\,,
\end{align}
where $p (b_1,v_1,b,v)$ is the quadrivariate distribution of $\delta(s_1)$, $\delta'(s_1)$, $\delta(s)$ and $\delta'(s)$, and $b_1\equiv b(s_1)=b$. It is straightforward to include a moving barrier, simply inserting $b'$ and $b_1'$ where needed (\`a la equation~\ref{fmeanv}).

Trajectories crossing more than once would now be overcounted: for instance, a single walk crossing at $s_1$ and $s_2$ would be removed twice by this procedure, and needs to be reintroduced. This would call for an additional correction, for walks crossing twice or more, containing $p(b_1,v_1,b_2,v_2,b,v)$ and integrals over $s_1$ and $s_2$, and so on. However, trajectories with more zigzags will be even more suppressed, making an expansion in the number of crossings meaningful in the sense of perturbation theory at small $s$.

Similarly to \eqn{fstheta} for the leading order term, the first subleading correction can also be written in a more evocative way as
\begin{equation}
  \bigg\langle{\bigg[\frac{\mathrm{d}}{\mathrm{d}s} \vartheta(\delta-b)\bigg]
  \vartheta(\delta'-b')
  \bigg[\frac{\mathrm{d}}{\mathrm{d}s_1} \vartheta(\delta_1-b_1)\bigg]
  \vartheta(\delta_1'-b_1')}\bigg\rangle\,,
\end{equation}
and the same pattern holds for higher order corrections.  A rigorous derivation of this expansion from a path integral expression is carried out in Appendix \ref{sec:corrections}.  However, for most cosmological applications, \eqn{fmnv} is sufficiently accurate:  \cite{ms12} showed that deviations exceed the percent level only for $s\gtrsim1$, so one does not even need the second term of \eqn{eq:corrections}.

\subsection{Gaussian or not?}

The logic above holds in full generality, regardless of the shape of the distribution: the completely correlated non-Gaussian walks have a modified $s$ dependence, but the first crossing distribution in this limit is still given by \eqn{fPS}, and this limit will still be a good approximation as $s\to 0$.  At larger $s$, an expansion in the number of previous upcrossings is still sensible, where constraining the slope of the walk is the most natural and efficient way of ensuring it is upcrossing.  So, \eqn{fmnv} should remain a good approximation to $f(s)$ until $s$ values where walks which can have previously upcrossed more than once dominate, at which point the next terms in the program (outlined in Section~\ref{multiple}) will become important.

That said, there is one sense in which the non-Gaussian case is more complicated.  For a Gaussian field, the probability distribution of $\delta$ on scale $s$ only depends on the ratio $\delta/\sqrt{s}$.  This makes 
\begin{equation}
  f_\mathrm{CC}(s) =
  - \left(\frac{\mathrm{d}}{\mathrm{d}s} \frac{b}{\sqrt{s}}\right)
  p\left(b/\sqrt{s}\right) \,,
\label{fcc}
\end{equation}
where for ease of notation we have not written the scale dependence of $b(s)$ explicitly. If the barrier is constant, $b=\delta_c$, then $\delta_c/\sqrt{s}$ is usually called $\nu$ and one finds $sf(s) = \nu\, p(\nu)/2$: the final factor of 1/2 is the reason Press \& Schechter multiplied by 2 so many years ago.  But notice that, in this limit, the first crossing distribution is very simply related to the shape of the pdf.  

The same would also apply to the non-Gaussian case, {\em provided} that the distribution of $\delta$ is indeed a function of $\delta/\sqrt{s}$ only. This is rarely the case, and in general one has
\begin{equation}
  f_\mathrm{CC}(s) = \left(\avg{\!v|b}-b'\right) p(b) \,,
\label{fcc2}
\end{equation}
where $\avg{\!v|b}$ is the mean of $p(v|b)$. For a non-Gaussian process the mean is no longer $b/2s$, unless the distribution depends on $s$ only through $\delta/\sqrt{s}$.  However, as we will see it becomes a reasonable approximation at small $s$.

%%%%%%%%%%%%%%%%%%%%%%%%%%%%%%%%%%%%%%%%%%%%%%%%%%%%%%%%%%%%%%%%%%

\section{Explicit calculation}

\label{sec:main}

%%%%%%%%%%%%%%%%%%%%%%%%%%%%%%%%%%%%%%%%%%%%%%%%%%%%%%%%%%%%%%%%%%

In what follows, it will be convenient to use the rescaled stochastic quantities
\begin{equation}
 \Delta\equiv\frac{\delta}{\sqrt{s}},\quad
 \Delta' \equiv \frac{\dd\Delta}{\dd s}\quad  {\rm and}\quad 
 \xi \equiv -\frac{\Delta'}{\sqrt{\avg{\Delta'^2}}}
     \equiv -2\Gamma s\,\Delta'
\end{equation}
where  $\Gamma$, defined by $(2\Gamma s)^2\equiv 1/\avg{\!\Delta'^2\!}$ is a weak function of $s$ \citep[e.g.][]{ms12}.  Notice that
\begin{equation}
 \avg{\Delta^2}=\avg{\xi^2}=1 \qquad {\rm and}\qquad 
 \avg{\Delta\,\xi}=0\,;
\end{equation}
i.e., $\Delta$ and $\xi$ are uncorrelated (although in general not independent) random variables. 
Similarly, we will work with
\begin{equation}
  B(s)\equiv \frac{b(s)}{\sqrt{s}}\qquad {\rm and}\qquad
  X \equiv -\frac{\dd B/\dd s}{\sqrt{\avg{\Delta'^2}}}
  = -2\Gamma s\, B',
\end{equation}
where $B' \equiv \dd B/\dd s$.
The sign of $X$ is chosen so that a typical barrier has $X>0$, since $b(s)$ for most problems of current interest does not vary much with $s$, and thus $B'<0$.  Since we are enforcing $\delta=b$, in these rescaled variables $f_\up$ reads 
\begin{equation}
%  f(s) \simeq 
  f_\up(s) = -B'p(B)\int_{-\infty}^X \!\!\dd\xi \,
  \bigg(1-\frac{\xi}{X}\bigg)\, p (\xi|B) \,.
\label{freduced}
\end{equation}
Equivalently, in the notation of \eqn{fstheta} one has
\begin{equation}
\label{fderiv}
%  f(s) \simeq 
  f_\up(s) = \frac{\mathrm{d}}{\mathrm{d}s} 
  \bigg[ \int_{B(s)}^{\infty} \!\!\!\dd \Delta
  \int_{-\infty}^{X(s_1)} \!\!\!\!\!\dd \xi \,p(\Delta,\xi;s,s_1)
 \bigg]_{s_1=s}\,,
\end{equation}
where $s_1$ must be set equal to $s$ after taking the derivative. Since $X\sim B$, we see explicitly that  we recover $f_\PS(s)$ in the large mass limit, when $X\gg1$.

%%%%%%%%%%%%%%%%%%%%%%%%%%%%%%%%%%%%%%%%%%%%%%%%%%%%%%%%%%%%%%%%%%

\subsection{Summary of the Gaussian result}

\label{sec:Gaussian}

%%%%%%%%%%%%%%%%%%%%%%%%%%%%%%%%%%%%%%%%%%%%%%%%%%%%%%%%%%%%%%%%%%

If $\delta$ is a Gaussian process, then the joint distribution of $\Delta$ and $\xi$ is particularly simple because $\avg{\!\Delta\,\xi\!}=0$.  When $\Delta=B$, then 
\begin{equation}
  p_\G(B,\xi) = p_\G(B)\,p_\G(\xi)
              = \frac{{\rm e}^{-B^2/2}}{\sqrt{2\pi}}\,
                \frac{{\rm e}^{-\xi^2/2}}{\sqrt{2\pi}}\,.
\label{pgauss}
\end{equation}
Inserting this in \eqn{freduced} shows that $f(s)$ will be proportional to $-B'p_\G(B)$, which, for a Gaussian process is just $f_\mathrm{CC}(s)$ times a correction factor that is a function of $X$ alone. Performing the integral yields
\begin{equation}
  f_\up(s) = -B'\,\frac{{\rm e}^{-B^2/2}}{\sqrt{2\pi}}
  \left[\frac{1 + {\rm erf}(X/\sqrt{2})}{2} 
       + \frac{{\rm e}^{-X^2/2}}{\sqrt{2\pi}X}\right].
\label{sfs}
\end{equation}
This reduces to \eqn{fcc} -- and therefore to $f_\PS(s)$ -- for $X\gg 1$ (the first term in the square brackets tends to unity while the second one is exponentially suppressed).  

For a wide variety of smoothing filters, power-spectra and barrier shapes, $f_\up$ is substantially more accurate than $f_\PS$, and remains accurate down to scales ($X\simeq 1$) on which a relevant fraction of the walks have negative slopes \citep{ms12}.  However, it cannot be accurate to arbitrarily small scales since, for constant barriers, the integral of $f_\up(s)$ over all $s$ diverges (altough its accuracy increases for moving barriers).  This is, of course, related to the fact that multiple upcrossings of the barrier become important as $s$ increases.  Appendix~\ref{sec:corrections} describes how to account for these, but since we have not found a similarly simple analytic expression for the resulting $f(s)$ (but see \citealt{ms13b} for an excellent and efficient numerical approximation), and the range over which \eqn{sfs} is accurate covers most of the range which is of interest in cosmology, we will continue with this simpler case.

Before moving on, we think the special case of Gaussian walks with correlated steps crossing a constant barrier deserves further comment.  Once one accounts for differences in notation and presentation, our \eqn{sfs} turns out to be the same as equation~(3.14a) of \cite{bcek91} for the first crossing distribution of a barrier of constant height.  The origin of this agreement is that the expression within angle brackets in our \eqn{fstheta} is the same as their equation~(3.12a). (The same is true for our equation \ref{simplestcase} and their A3.)  Unfortunately, their Figure~9 emphasizes the fact that ignoring multiple crossings is a bad approximation in the low mass limit.  This led them, and the rest of the field since, to dismiss the approximation in which one ignores multiple upcrossings, and to focus instead on what appeared to be a more tractable problem (in which one ignores correlations between steps).  However, as emphasized by \citet{ms12} in their more general analysis, the upcrossing approximation is in fact rather good in the high mass limit, and  works well for arbitrary (i.e. not just constant) barriers.  We now show how it can be generalized to arbitrary fields.

%%%%%%%%%%%%%%%%%%%%%%%%%%%%%%%%%%%%%%%%%%%%%%%%%%%%%%%%%%%%%%%%%%

\subsection{Generic non-Gaussian case}
\label{sec:generic}

%%%%%%%%%%%%%%%%%%%%%%%%%%%%%%%%%%%%%%%%%%%%%%%%%%%%%%%%%%%%%%%%%%

The joint probability distribution of a generic stochastic process can always be written as an asymptotic expansion in Hermite polynomials around the Gaussian distribution obtained from the second moments. Since $\Delta$ and $\xi$ are uncorrelated, their distribution is 
\begin{equation}
\label{pjointNG}
  p(B,\xi) = \sum_{n,k} 
  \frac{\avg{\!H_n(\hat\Delta)H_k(\hat\xi)\!}}{n!\,k!}
             H_n(B)H_k(\xi)\,p_\G(B,\xi)\,,
\end{equation}
where we have used hats to distinguish the stochastic quantities from the continuous variables of the probability distribution, and $H_n(x)\equiv \exp(x^2/2)(-\dd/\dd x)^n\exp(-x^2/2)$. This expression follows from the fact that the Hermite polynomials form an orthogonal basis with respect to the Gaussian weight.  Rearranging the terms of the sum and factoring out $p(B)$ shows that 
\begin{equation}
\label{pcondNG}
  p(\xi|B) = \sum_{k}
  \frac{\avg{\!H_k(\hat\xi)|B}}{k!}%H_k(\xi)\,
  \left(\!-\frac{\dd}{\dd\xi}\right)^{\!k}\!p_\G(\xi)\,,
\end{equation}
where 
\begin{equation}
  \avg{\!H_k(\hat\xi)|B} = 
  \frac{\sum_n\avg{\!H_n(\hat\Delta)H_k(\hat\xi)\!}H_n(B)/n!}
  {\sum_n\avg{\!H_n(\hat\Delta)\!}H_n(B)/n!}\,
\label{condcorr}
\end{equation}
and $\avg{f(\hat\xi)|B}$ is the expectation value of $f(\hat\xi)$ given that $\hat\Delta=B$, i.e.~the one computed from $p(\xi|B)$.

This expression must be inserted into \eqn{freduced}, and integrated over $\xi$. The $k=0$ term is just $p_\G(\xi)$, and gives the same as \eqn{sfs}, with $p_\G(B)$ replaced by $p(B)$. 
The $k=1$ term can be integrated by parts and gives $-[\avg{\!\hat\xi|B}p(B)/(2\Gamma s)][1+\mathrm{erf}(X/\sqrt{2})]/2$. However, $\avg{\!\hat\xi|B}p(B)$ is $p_\G(B)$ times the numerator of \eqn{condcorr} with $k=1$, for which one has
%\begin{equation}
$  \avg{\!H_n(\hat\Delta)\hat\xi\!} =   
  -2\Gamma s\avg{\!H_{n+1}(\hat\Delta)\!}'\!/(n+1)$.
%\end{equation}
Furthermore, $H_n(B)p_\G(B)=\int_B^\infty\!\dd\Delta \,H_{n+1}(\Delta)p_\G(\Delta)$, so that 
\begin{equation}
  p(B)\avg{\!\hat\xi|B\!}
  = -2\Gamma s\int_B^\infty\!\dd\Delta \,[\pd p(\Delta)/\pd s]\,.
\end{equation}
Finally, for $k\geq2$ all terms can be integrated by parts %The following ones can be integrated by parts, and they pick up a factor of $(B'/X)(-\pd/\pd X)^{k-1}[1+\mathrm{erf}(X/\sqrt{2})]/2$, that for $k\geq2$ becomes 
and pick up a factor of $H_{k-2}(X)p_\G(X)/(2s\Gamma)$.
Putting all the contributions together yields
\begin{align}
  f_{\up}(s) &=  \bigg[\frac{\dd}{\dd s} \!
  \int_{B(s)}^{\infty} \!\!\!\dd\Delta\,p(\Delta) \bigg] 
  \frac{1 + {\rm erf}(X/\sqrt{2})}{2} \notag \\
  &-B'p(B) \frac{e^{-X^2/2}}{X\sqrt{2\pi}}
  \bigg[1+\!\sum_{k=2}^\infty \!\frac{\avg{\!H_k(\hat\xi)|B\!}}{k!}H_{k-2}(X)\bigg]
\label{eq:fNGeval}
\end{align}
as the non-Gaussian generalization of \eqn{sfs}.  Equation~(\ref{eq:fNGeval}), and the closely related \eqn{fNGrenorm} below, are the main results of this paper.  The connection is most clearly seen by supposing that $p(\delta;s)$ is a function of $\delta/\sqrt{s}$ alone, in which case the derivative of the integral in the first line, which we could have written as $f_\PS(s)$, becomes $-B'p(B)$.  Then \eqn{eq:fNGeval} -- apart from the polynomial corrections in the square brackets of the second line, which we suppose small -- becomes \eqn{sfs}, except that here $p(B)$ is the full non-Gaussian distribution.  In general, of course $p(\delta;s)$ will not be a function of $\delta/\sqrt{s}$ only; the associated departure from self-similarity will introduce an additional term, and this is why \eqn{eq:fNGeval} involves an explicit derivative with respect to $s$.

In the large mass regime, when $X\gg1$, the entire second line of \eqn{eq:fNGeval} is suppressed with respect to the first one, as it was in the Gaussian case, and the first crossing rate reduces to \eqn{fPS}.  The Hermite polynomial corrections in the second line of \eqn{eq:fNGeval} would singularly blow up at large $X$, but are in this regime suppressed by the exponential cutoff $\exp(-X^2/2)$. They have a chance of becoming non-negligible only at larger $s$, when $X\sim 1$ and the exponential suppression is no longer effective. In this regime, however, the perturbative treatment of these non-Gaussian corrections is fully under control.

%%%%%%%%%%%%%%%%%%%%%%%%%%%%%%%%%%%%%%%%%%%%%%%%%%%%%%%%%%%%%%%%%%
\subsection{Large non-Gaussianity}
\label{largeNG}
%%%%%%%%%%%%%%%%%%%%%%%%%%%%%%%%%%%%%%%%%%%%%%%%%%%%%%%%%%%%%%%%%%

Equation~(\ref{eq:fNGeval}) is particularly well suited for a case -- like primordial non-Gaussianity -- where all the non-linear moments $\avg{\!\Delta^n\!}_c$ are small. For such a process, deviations from linearity are relevant only when $B$ grows large, as described in the previous Section, while for $B\sim1$ all corrections are perturbative.  However, one may worry that if the non-Gaussian moments of the distribution are large, corrections from the series in \eqn{eq:fNGeval} may become relevant at large $s$.  This happens because the Edgeworth expansion of $p(\xi|B)$ around $p_\G(\xi)$, that is \eqn{pcondNG}, is badly behaved if the conditional mean and variance significantly differ from 0 and 1, and cannot be truncated at any order.

%$\mu\equiv\avg{\xi|B}$ 
%$\Sigma\equiv 1+\avg{\!H_2(\xi)|B}-\avg{\xi|B}^2$ 

Still, regardless of the size of the non-Gaussian corrections, \eqn{fmnv} may always be written -- exactly -- as
\begin{equation}
  f_\up(s) = f_\PS(s) \int_{b'}^{\infty}\!\! \dd v
  \bigg(1-\frac{\mu-v}{\mu-b'}\bigg) p(v|b)\,,
\end{equation}
with $\mu\equiv\avg{\!v|b}$ and $f_\PS(s)$ given by \eqn{fcc2}.
If the mean and variance of $p(v|b)$ significantly differ from their Gaussian values $b/2s$ and $1/(4\Gamma^2s)$, then it is more appropriate to write $p(v|b)$ as an Edgeworth series around a Gaussian with mean and variance given by the true renormalized non-Gaussian values.  One can then define a new rescaled variable $\tilde\xi\equiv(\mu-v)/\sqrt{\avg{\!v^2|b}-\mu^2}$, and following exactly the same steps as in Section \ref{sec:generic}, one gets
\begin{align}
  f_\up(s) = f_\PS(s)\,&\bigg[
%  f(s) &= \frac{\mu-b'}{\sqrt{s}}p(B)\bigg[
%  \bigg[\frac{\dd}{\dd s} \!
%  \int_{B(s)}^{\infty} \!\!\!\dd\Delta\,p(\Delta) \bigg] 
  \frac{1 + {\rm erf}(Y/\sqrt{2})}{2}
  + \frac{e^{-Y^2/2}}{\sqrt{2\pi}Y}\notag \\
  &
  + \sum_{k=3}^\infty \frac{\avg{\!H_k(\tilde\xi)|B\!}}{k!}H_{k-2}(Y)
  \frac{e^{-Y^2/2}}{\sqrt{2\pi}Y}\bigg],
\label{fNGrenorm}
\end{align}
where 
\begin{equation}
  Y \equiv \frac{\mu-b'}{\sqrt{\avg{\!v^2|b}-\mu^2}} =
%  - \frac{2\Gamma s B' + \avg{\xi|B}}{
%    \sqrt{\avg{\!\xi^2|B}-\avg{\xi|B}^2}}\,.
  \frac{ \avg{\Delta'|B}-B'}{
    \sqrt{\avg{\!\Delta'^2|B}-\avg{\Delta'|B}^2}}
\end{equation}
accounts for the non-Gaussian corrections to the mean and variance of the conditional distribution in a non-perturbative way.

This expression is very similar to \eqn{eq:fNGeval}, and, like it, is the main result of this paper.  While it has the same large scale ($s\ll 1$) limit as before, $f_\PS(s)$, the entire expression here is proportional to $f_\PS(s)$, whereas, for \eqn{eq:fNGeval} this only happened if $\avg{\Delta'|B}=0$ (for which $f_\PS(s)=-B'p(B)$).
%as , but depends on the renormalized mean and variance of the non-Gaussian $p(\Delta'|B)$ in a non-perturbative way. 
Moreover, now the series starts at $k=3$, as all terms like $\avg{\!\Delta'\Delta^n\!}_cB^n$ and $\avg{\!\Delta'^2\Delta^n\!}_cB^n$, which can become important at small scales if the moments are large, are included in $Y$.  In this respect, \eqn{fNGrenorm} is a more efficient expansion than is \eqn{eq:fNGeval}.  Of course, one should still worry about further corrections from all the non-linear moments like $\avg{\!\Delta'^3\Delta^n\!}_c$. If these were of the same order (as they in principle are), one should keep all the terms of the infinite series in \eqn{fNGrenorm}. If $p(\xi|B)$ is known one can compute them by writing
\begin{equation}
  \sum_{k=3}^\infty \frac{\avg{\!H_k(\tilde\xi)|B\!}}{k!}H_{k-2}(Y)
  \frac{e^{-Y^2/2}}{\sqrt{2\pi}}%p_\G(Y)
  = \int_{-\infty}^Y\!\!\!\!\dd y \,C(y)\,,
\end{equation}
where $C(y)\equiv\int_{-\infty}^y\!\dd\tilde\xi \,p(\tilde\xi|B)-[1+\mathrm{erf}(y/\sqrt{2})]/2$ is the difference between the cumulative distributions of $p(\tilde\xi|B)$ and $p_\G(\tilde\xi)$. However, we will see shortly that there is a broad and fairly generic class of models for which all these corrections vanish identically.

Before moving on, we note that a remarkable feature of both equations~(\ref{eq:fNGeval}) and~(\ref{fNGrenorm}) is that, although we started from the non-Gaussian bivariate distribution $p(b,v;s)$, the upcrossing distribution $f_\up(s)$ can be expressed in terms of a function modulating the univariate non-Gaussian distribution $p(b;s)$.  Therefore, for the scales where $f(s)\simeq f_\up(s)$, we have managed to disentangle the problem of the first crossing of the barrier from that of the evaluation of the probability of the walks, which we deal with next.

%%%%%%%%%%%%%%%%%%%%%%%%%%%%%%%%%%%%%%%%%%%%%%%%%%%%%%%%%%%%%%%%%%
\subsection{Non-Gaussian case: large mass limit}
\label{sec:large}
%%%%%%%%%%%%%%%%%%%%%%%%%%%%%%%%%%%%%%%%%%%%%%%%%%%%%%%%%%%%%%%%%%

We have already argued that in the large mass limit the formal expression for the first crossing rate coincides with \eqn{fPS}.
If $p(B)$ is known, then $f(s)$ can be written as
%To see this explicitly, differentiate with respect to $s$ to get 
\begin{equation}
\label{largemass}
  f_\PS(s) = \bigg[-B' +
  \sum_{n=3}^{+\infty}
  \frac{\avg{\!\Delta^n\!}_c'}{n!}
  \left(\!-\frac{\pd}{\pd B}\right)^{\!n-1}\bigg]
  p(B;s)\;,%\frac{e^{\D_0}e^{-\beta^2/2}}{\sqrt{2\pi}}\;,
\end{equation}
where the infinite sum is the (integral of the) Kramers-Moyal expansion for $\pd p/\pd s$. This sum is what gives $\avg{\Delta'|B}p(B)$.
Often, however, the single point $p(B)$ is not known:  only its moments are.  The crucial point, therefore, is to estimate $p(B)$ from its moments.  This can be written as
\begin{equation}
 \label{pBexpW}
  p(B;s)=\frac{{\rm e}^{W(B;s)}}{\sqrt{2\pi}}\,,
\end{equation}
where the full expression of $W(B;s)$ in terms of the moments is given in Appendix \ref{sec:pdf} as a series of modified Hermite polynomials.

The function $W$, which corresponds to the logarithm of the Edgeworth expansion of $p(B;s)$, has a straightforward interpretation in terms of connected Feynman diagrams constructed out of the connected moments of the distribution, and better convergence properties that the Edgeworth expansion itself.
We also show in the Appendix that the large mass limit ($B \gg 1$) of $W(B;s)$ is obtained by keeping only the highest order term of each polynomial, which in diagrammatic language corresponds to discarding diagrams with loops. This approximation is also what one would get by doing the analysis in Fourier space and transforming back to real space by means of a stationary phase approximation.  In this regime, the infinite series of polynomials turns into a simpler infinite power series, whose first terms are
\begin{equation}
\label{W}
  W(B;s) \simeq -\frac{B^2}{2} +
   \frac{\avg{\!\Delta^3\!}_c}{3!}B^3
  +\frac{\avg{\!\Delta^4\!}_c - 3\avg{\!\Delta^3\!}_c^2}{4!}B^4 
  + \dots\,.
\end{equation}
In the same spirit, we can approximate the $n$-th derivative as $(\pd/\pd B)^np(B)\simeq (\pd W/\pd B)^np(B)$, since higher derivatives of $W(B;s)$ also correspond to loop diagrams and are subleading.

The consistency of the truncation of $W(B;s)$ is a delicate subject. Clearly, as $B$ becomes large one should keep adding more and more terms to Eq.~\eqref{W}, especially if the non-Gaussian moments are large, and a true $B\to \infty$ limit would necessarily require resumming the whole series. Fortunately, the range of values of interest for Cosmology (where $B$ increases both with mass and redshift) is not so extreme, since primordial non-Gaussianities are fairly small (we quantify this shortly).

Truncating the Kramers-Moyal series in Eq.~\eqref{largemass} is on the other hand less dangerous. The reduced moments typically tend to constants on large scales, and the appearance of their derivatives in the coefficients of the series introduces additional suppression. Furthermore, this series does not sit in an exponential, so errors in the truncation are potentially less harmful. Even for the $n=3$ term of the series, keeping just the leading term of $\pd W/\pd B$ gives a $\mathcal{O}(1)$ result (or less, given the additional suppression due to the scale derivative). Within the range of parameters outlined above, a fair approximation for the first crossing rate is thus
\begin{equation}
 \label{fsW}
  f(s)\simeq\bigg(\!\!-B' +
  \frac{\avg{\!\Delta^3\!}_c'}{3!} B^2 \bigg)
  \frac{{\rm e}^{W(B;s)}}{\sqrt{2\pi}}\,,
\end{equation}
with $W(B;s)$ given by Eq.~\eqref{W}.  

In many cases $\avg{\!\Delta^3\!}_c$ (and more generally $\avg{\!\Delta^n\!}_c$) is only weakly scale dependent. If we can drop the $\avg{\!\Delta^3\!}_c'$ term, then \eqn{fsW} simplifies even further, reducing to \eqn{fcc}. In this regime, $f(s)$ is just $-B'p(B)$, so that the ratio of the non-Gaussian result to the Gaussian one simply corresponds to the ratio of the pdf's:
\begin{equation}
 \label{fNGratio}
 \frac{f_{\rm NG}(s)}{f_{\rm Gauss}(s)}\to \frac{{\rm e}^{W(B)}}{{\rm e}^{-B^2/2}} 
  = \exp(\avg{\!\Delta^3\!}_c B^3/3! + \ldots)\,.
\end{equation}
Evidently, the difference from the Gaussian case will only be apparent if $\avg{\!\Delta^3\!}_c B^3$ is large; it will be more obvious at the large $B$, i.e. in the large mass tail.  We will return to this limit in the next section.

%%%%%%%%%%%%%%%%%%%%%%%%%%%%%%%%%%%%%%%%%%%%%%%%%%%%%%%%%%%%%%%%%%
\subsection{Testing the model: non-linear transformations of Gaussian variables}
\label{test}
%%%%%%%%%%%%%%%%%%%%%%%%%%%%%%%%%%%%%%%%%%%%%%%%%%%%%%%%%%%%%%%%%%

We now test our formalism in a few simple cases that also admit an analytical treatment.  A common way to obtain a non-Gaussian distribution for $\delta$ is to apply a non-linear transformation to a Gaussian variable $\delta_\L$. Doing this on every scale $s_\L\equiv\avg{\delta_\L^2}$ maps the Gaussian walk $\delta_\L(s_\L)$ into a non-Gaussian walk $\delta(s)$.  As a result, the barrier $b$ for the non-Gaussian variable maps into an effective barrier $b_\L$ that $\delta_\L$ must cross. A similar mapping links the reduced variables $\Delta$ and $B$ to their Gaussian counterparts $\Delta_\L\equiv\delta_\L/\sqrt{s_\L}$ and $\BL$.  The non-Gaussian first crossing distribution $f(s)$ is then related to the Gaussian $f(s_\L)$ for the barrier $b_\L$, as 
\begin{equation}
 \label{fsfsL}
 f(s) = \frac{\dd s_\L}{\dd s}\,f(s_\L).
\end{equation}
Hence, on the scales where $f(s_\L)\simeq f_\up(s_\L)$, we can simply 
%where $f(s_\L)$ can be obtained from 
insert $\BL$, $\dd\BL/\dd s_\L$ and $\Gamma_\L^2\equiv 1/[4s_\L^2\avg{\!(\dd\DL/\dd s_\L)^2\!}] $ in \eqn{sfs} to obtain a good approximation to $f(s)$.

\subsubsection{General formalism}

In general, an arbitrary transformation $\Delta=F(\Delta_L,s_\L)$ will also have an explicit dependence on $s_\L$. Differentiating w.r.t. $s$ gives
\begin{equation}
  \Delta'=\frac{\dd s_\L}{\dd s}
  \bigg[\frac{\pd F}{\pd \DL} \DL'
  + \frac{\pd F}{\pd s_\L}\bigg],
\label{transv}
\end{equation}
where $\DL'\equiv\dd \DL/\dd s_\L$.
This means that the mean and variance of the conditional distribution $p(\Delta'|B)$ are %(\dd s_\L/\dd s)\pd F(\BL,s_\L)/\pd s_\L
\begin{align}
  \avg{\Delta'|B} = \frac{\dd s_\L}{\dd s}
  \frac{\pd F(\BL,s_\L)}{\pd s_\L}
\end{align}
and
\begin{equation}
  \avg{\Delta'^2|B}-\avg{\Delta'|B}^2 = 
  \bigg[\frac{\dd s_\L}{\dd s} \frac{\pd F(\BL,s_\L)}{\pd \BL}\bigg]^2
  \avg{\DL'^2}\,.
\end{equation}
In addition, at fixed $B$, the relation between $\Delta'$ and $\DL'$ is linear, so all higher moments of $p(\Delta'|B)$ vanish identically, and hence, the entire second line of \eqn{fNGrenorm} vanishes as well.  

If we assume for simplicity that the transformation is monotonic (although this requirement can be easily relaxed), then $B_\L=F^{-1}(B)$ is unique and $p(B)=(\dd \BL/\dd B)p_\G(\BL)$. Therefore, at small $s$, $f(s)$ is well approximated by
\begin{equation}
  f_\PS(s) =
%  \frac{\dd}{\dd s}\int_b^{\infty}\!\!\dd\Delta\, p(\Delta)
  \left(\avg{\!\Delta'|B\!}-B'\right)p(B)
  =-\frac{\dd s_\L}{\dd s}\frac{\dd B_\L}{\dd s_\L}
  \frac{e^{-B_\L^2/2}}{\sqrt{2\pi}}\,.
\label{fccNL}
\end{equation}
%\begin{equation}
%  \int_{-\infty}^{b_\L}\!\!\!\dd\delta_\L \,p_\G(\delta_\L;s)=
%  \int_{-\infty}^{b}\!\!\!\dd\delta \,p(\delta;s)\,,
%\end{equation}
At intermediate $s$ one need only use the first line of \eqn{fNGrenorm}, as each term in the second line equals zero, with
\begin{equation}
  Y = \frac{ \pd F/\pd s_\L - (\dd s/\dd s_\L)B'}{|\pd F/\pd \BL|\sqrt{\avg{\DL'^2}}}
  = - \frac{\dd B_\L/\dd s_\L}{\sqrt{\avg{\DL'^2}}} 
  \equiv X_\L
\label{effX}
\end{equation}
as the counterpart of $X$ for the Gaussian walks.
This confirms the intuition that the first crossing distribution for these processes can be obtained from that of the Gaussian walks crossing the effective barrier $\BL$. 

All that is left to do is to relate $\avg{\Delta'^2}=1/(2\Gamma s)^2$ to its Gaussian counterpart $\avg{\DL'^2}$.  Taking the variance of \eqn{transv} gives
\begin{equation}
  \bigg(\frac{\dd s/\dd s_\L}{2\Gamma s}\bigg)^{\!2} =
  \avg{\bigg(\frac{\pd F}{\pd \DL}\bigg)^{\!2}} \avg{\DL'^2}
  + \avg{\bigg(\frac{\pd F}{\pd s_\L}\bigg)^{\!2}}\,,
\end{equation}
which completes the link. Notice that even after $\Delta=F(\DL)$ is given, both $\dd s/\dd s_\L$ and $\avg{\DL'^2}$ (since $\DL'$ is independent of $\DL$ and $\Delta$) are essentially free parameters, that can be used for additional tuning of the transformation. This result (i.e., \eqn{fNGrenorm} without the second line) still neglects multiple crossings, but the treatment of non-Gaussianity is otherwise exact.

\subsubsection{A fully worked example}
Consider the first crossing distribution of a constant barrier $b=\delta_c$ by Lognormal walks given by the exponential map
\begin{equation}
   1+\delta/\delta_c=\exp(\delta_\L/\delta_c-s_\L/2\delta_c^2)\,.
\label{logdef}
\end{equation}
%$\delta/\delta_c \equiv \rho/\avg{\rho}-1$ with $\rho\equiv\exp(\delta_\L/\delta_c)$. % For generic $\delta_*$ and $\alpha$, 
%This transformation has $\avg{\delta}=0$, $\avg{\rho} = \exp(s_\L/2\delta_c^2)$ and $s = \delta_c^2[\exp(s_\L/\delta_c^2)-1]$. 
Since  $\avg{(1+\delta/\delta_c)^n}=\exp[n(n-1)s_\L/2\delta_c^2]$, these walks have $\avg{\delta}=0$ and
\begin{equation}
  1+\frac{s}{\delta_c^2}%\equiv 1+\frac{1}{B^2}
  =\exp(s_\L/\delta_c^2) =\frac{\dd s}{\dd s_\L}
  \equiv\Sigma \,.
\end{equation}
Inverting these relations, we get $\delta_\L=\delta_c\ln(1+\delta/\delta_c) + s_\L/2\delta_c$ and $s_\L=\delta_c^2 \ln(1+s/\delta_c^2)$.
The distribution of $\delta$ is therefore Lognormal, with
\begin{equation}
%  p(\Delta)=\sqrt{s}
  p(\delta) = \frac{p_\G(\delta_\L)}{1+\delta/\delta_c} =
  \frac{e^{-\delta_\L^2/2s_\L}}{(1+\delta/\delta_c)\sqrt{2\pi s_\L}};
\label{LNpdf}
\end{equation}
in the next section we use this transformation to mimic the distribution of the final (rather than primordial) density field.  
%Here, we are more interested in using it as a toy-model of non-Gaussian initial conditions.  
%with $p(\delta)=p_\G(\delta_\L)/(1/\alpha+\delta/\delta_*)$.

%We list here some useful expressions to compute the first crossing 
%distribution of Lognormal walks given by the transformation
%\begin{equation}
%   1+\delta/\delta_c=\exp(\delta_\L/\delta_c-s_\L/2\delta_c^2),
%\label{logdef}
%\end{equation}
%crossing a constant barrier $\delta_c$. 

A constant barrier $\delta_c$ for the Lognormal walks becomes a linearly increasing barrier 
 $b_\L = \delta_c\ln(2) + s_\L/2\delta_c$ 
for the Gaussian walks.  This illustrates nicely the rather general result that a constant barrier for the non-Gaussian walks translates to a moving effective barrier for the underlying Gaussian walks.  %Equivalently, the rescaled effective barrier 
%In this example we define $\rho\equiv\exp(\delta_\L/\delta_c)$ and $1+\delta/\delta_c \equiv \rho/\avg{\rho}$, for which $\avg{\delta}=0$, $\avg{\rho} = \exp(\sigma^2_\L/2\delta_c^2)$, and $1+s/\delta_c^2 = \exp(\sigma^2_\L/\delta_c^2)$.  
%
% On the other hand, setting $\delta_*=\sqrt{s_\L}$ and $\alpha^2=e-1$ one gets $s=s_\L$, $\BL=\log(1+\sqrt{e-1}B)+1/2$ and $\avg{\DL'^2}= (e-1)\avg{\Delta'^2}/e$.
%
%and $\pd F/\pd s_\L=0$, so that $\avg{\Delta'|B}=0$ like in the Gaussian case. However, with these coefficients one also has $\pd F/\pd\DL=\Delta+1/\sqrt{e-1}$ and $\avg{\Delta'^2}= e \avg{\DL'^2}/(e-1)$, which means that %but $\avg{\Delta'^2|B}= (B+1/\sqrt{e-1})^2 \avg{\DL'^2}$: the variance of $p(\Delta'|B)$ still depends on $B$, unlike in the Gaussian case. 
%
%The parameter describing the small scale modulation of $f(s)$ is in this case
%\begin{equation}
%  Y = - \frac{2\Gamma s B'\sqrt{e}}{1+\sqrt{e-1}B}\,,
%\end{equation}
%and has to be compared with $X=-2\Gamma s B'$ that we would use if we were evaluating $f(s)$ with \eqn{eq:fNGeval}, the whole series in the second line of which therefore resums into the simple rescaling $X\to Y$. 
%
%The first crossing distribution for these Lognormal walks crossing a barrier of constant height is the same as for Gaussian walks crossing $b_\L = \ln(2)\delta_c + \sigma^2_\L/2\delta_c$, %= \delta_c\ln(2\sqrt{\Sigma})$, upto a factor $\dd\sigma_\L^2/\dd s = (1+s/\delta_c^2)^{-1}$.
%
% and defining $\Sigma\equiv 1+s/\delta_c^2$, so that $b_\L = \delta_c\ln(2\sqrt{\Sigma})$, 
When $s\ll 1$, $f(s)$ is given by \eqn{fccNL} with   
\begin{equation}
  \BL = \frac{\delta_c}{\sqrt{s_\L}}
  \left(\ln 2 + \frac{s_\L}{2\delta_c^2}\right) 
  =\frac{\ln2}{\sqrt{\ln\Sigma}}+\frac{\sqrt{\ln\Sigma}}{2}
\end{equation}
and $\dd s_\L/\dd s = 1/\Sigma$. Working out the computation gives
\begin{align}
  f_\PS(s) %&= 
%  -\frac{1}{(1+s/\delta_c^2)}\frac{\dd (b_\L/\sqrt{s_\L})}{\dd s_\L}
%  \frac{e^{-b_\L^2/2s_\L}}{\sqrt{2\pi}}
%  \notag \\
  &= \bigg(\frac{\ln2}{\ln\Sigma}-\frac{1}{2}\bigg)
  \frac{2^{-[3+(\ln2/\ln\Sigma)]\!/2}}{\delta_c^2\,\Sigma^{9/8}\sqrt{2\pi\ln\Sigma}}\,.
%  \frac{\exp(-b^2_\L/2\sigma^2_\L)}{1+s/\delta_c^2}\,,
\end{align}
%where we have set $\Sigma\equiv (1+s/\delta_c^2)$.  
Conversely, for a constant barrier, $B'=-\delta_c/2s^{3/2}$ and 
\begin{equation}
  -B' p(B) = \frac{\delta_c}{2s}\, p(b)
  = \frac{\delta_c}{4s}\, p_\G(b_\L)
  = \frac{2^{-[3+(\ln2/\ln\Sigma)]\!/2}}{4s\Sigma^{1/8}\sqrt{2\pi\ln\Sigma}}\,,
\end{equation}
where the second equality comes from setting $\delta=\delta_c$ in \eqn{LNpdf}. This shows explicitly that $f_\PS(s) \neq -B' p(B)$.
%where in terms of $s$ one writes $ b_\L= \delta_c\ln(2\sqrt{1+s/\delta_c^2})$.

At intermediate $s$ one must use \eqn{fNGrenorm} instead (where the second line vanishes), inserting in it $Y$ from \eqn{effX}. That is, 
\begin{equation}
  Y = X_\L = -2\Gamma_\L s_\L \frac{\dd \BL}{\dd s_\L}
    = \Gamma_\L \Sigma \sqrt{\ln\Sigma}
      \bigg(\frac{\ln2}{\ln\Sigma}-\frac{1}{2}\bigg).
\end{equation}
This includes all non-Gaussian effects exactly, and will thus be different from using \eqn{eq:fNGeval} and ignoring the sum over Hermite polynomials in it, both because $-B' p(B)\ne f_\PS(s)$ and because $X\ne X_\L$.  To see this last point, however, we need the explicit relation between $\Gamma$ and $\Gamma_\L$.

Differentiating \eqn{logdef} w.r.t. $s$ one has
\begin{equation}
   v \equiv \frac{\dd\delta}{\dd s}
   = \bigg(v_\L-\frac{1}{2\delta_c}\bigg)
\exp\bigg(\frac{\delta_\L}{\delta_c}-\frac{3}{2}\frac{s_\L}{\delta_c^2}\bigg),
\end{equation}
where $v_\L\equiv\dd\delta_\L/\dd s_\L$, for which $\avg{v}=0$ and
\begin{equation}
   \avg{v^2}=\frac{\avg{v_\L^2}+1/4\delta_c^2}{1+s/\delta_c^2}\,,
%  \left(\avg{v_\L^2}+1/4\delta_c^2\right)\exp(-s_\L/\delta_c^2)
\end{equation}
or equivalently
\begin{equation}
   \Gamma^2 = \frac{1}{4s\avg{v^2}-1}
   = \frac{1+s/\delta_c^2}{4s\avg{v_\L^2}-1}\,;
\label{Gammalog}
\end{equation}
%using $\avg{v^2}=s\avg{\Delta'^2}+1/4s$ and 
using $4s_\L\avg{v^2_\L}=1+1/\Gamma_\L^2$ and
solving for $\Gamma_\L$ returns
%and since the barrier $b=\delta_c$ is constant one has 
%%$X\equiv-2\Gamma s (b/\sqrt{s})'=\Gamma\delta_c/\sqrt{s}$
%\begin{equation}
%   X\equiv -2\Gamma s B' = \Gamma\frac{\delta_c}{\sqrt{s}}=
%   \bigg[\frac{\delta_c^2/s+1}{4s\avg{v_\L^2}-1}\bigg]^{1/2}.
%\end{equation}
%
%The effective barrier $b_\L$ is obtained solving for $\delta_\L$ as a function of $\delta$ and then setting $\delta=b=\delta_c$. From this one has
%\begin{equation}
%   \BL = \frac{b_\L}{\sqrt{s_\L}} =
%   \frac{\delta_c}{\sqrt{s_\L}}\log 2 + \frac{\sqrt{s_\L}}{2\delta_c}.
%\end{equation}
%Moreover, in order to compute $Y\equiv -(\dd\BL/\dd s_\L)/\sqrt{\avg{\DL'^2}}$, \eqn{Gammalog} can be recast as
\begin{equation}
   \frac{1}{\Gamma_\L^2} = \frac{\ln\Sigma}{\Gamma^2}
   \bigg[1+\frac{\delta_c^2}{s}+\Gamma^2\bigg(\frac{\delta_c^2}{s}-\frac{1}{\ln\Sigma}\bigg)\bigg].
\end{equation}
This expression for $\Gamma_\L$ leads to
\begin{equation}
  Y%\equiv -\frac{\dd\BL/\dd s_\L}{\sqrt{\avg{\DL'^2}}}
%\bigg[\frac{\log 2}{\log(1+s/\delta_c^2)} - \frac{1}{2}\bigg]
%\frac{[\log2/\log(1+s/\delta)]-1/2}{\sqrt{1+(1+\Gamma^2)\delta_c^2/s 
%+[\Gamma^2/\log(1+s/\delta_c^2)]}}.
  = \frac{\Gamma\Sigma(\ln 2/\ln\Sigma-1/2)}{
  \sqrt{1+\delta_c^2/s +\Gamma^2(\delta_c^2/s -1/\ln\Sigma)}}\,,
\end{equation}
which clearly differs from $X=\Gamma\delta_c/\sqrt{s}$. Even for small $s$, when $\ln\Sigma\to s/\delta_c^2$, one still gets that $X_\L=Y\to X\ln2 $.

\begin{figure}
 \centering
 \includegraphics[width=0.9\hsize]{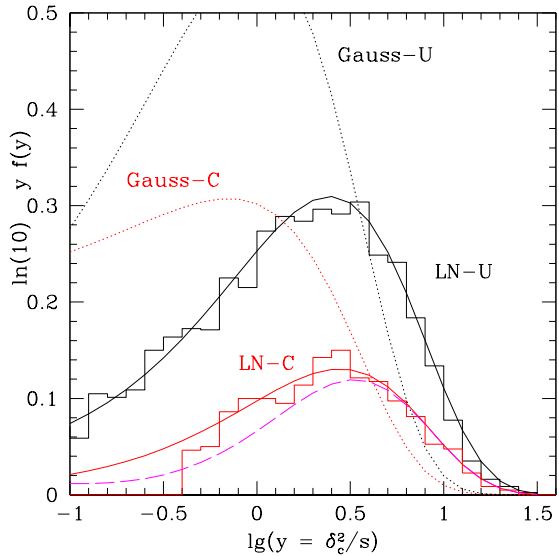}
 \caption{\label{lnWalks} First crossing distribution of a barrier of constant height $\delta_c$ by Gaussian (dotted curves) and Lognormal (histograms) walks having correlated and uncorrelated steps. The latter are obtained by applying the exponential map to the former. The correlations between steps come from Gaussian smoothing of a power spectrum with $P(k)\propto k^{-1}$.  The upper solid curve shows $sf(s)$ for Lognormal walks with uncorrelated steps, obtained via \eqn{fsfsL} by transforming the exact solution for Gaussian walks crossing a linearly increasing barrier $s_\L f(s_\L)$.  The lower solid curve shows \eqn{fNGrenorm} for Lognormal walks with correlated steps, equivalent to transforming the approximate solution to the Gaussian problem with linear barrier. The dashed curve shows \eqn{eq:fNGeval} with all Hermite correction terms set to zero.}
\end{figure}

To check the accuracy of our analysis we have computed numerically the first crossing distribution via Montecarlo realizations of Lognormal walks, obtained by applying the exponential map to Gaussian walks with both correlated and uncorrelated steps. The histograms in Figure~\ref{lnWalks} show the result for a barrier of constant height $\delta_c$ in these two cases. 
%which becomes a linearly increasing barrier for the Gaussian walks $\delta_\L$.
The Lognormal walks clearly have more first crossings at large $\delta_c^2/s \gg 1$, illustrating that we are working in the regime of large non-Gaussianity.  This is true for walks with correlated and uncorrelated steps, and is mainly a consequence of the fact that at large masses $s_\L\simeq s$ but $b_\L\simeq \delta_c\ln2 <\delta_c$, which makes reaching the barrier easier.

Although all other results in this paper concerned walks with correlated steps, we have also included uncorrelated steps here as a consistency check of our formalism. The first crossing distribution $f(s_\L)$ for the Gaussian walks crossing a linearly increasing barrier has a known exact solution (Sheth 1998), so $f(s)$ obtained through \eqn{fsfsL} is also exact. 
The upper solid curve, which provides an excellent description of the upper histogram, shows $sf(s)$ for Lognormal walks with uncorrelated steps, obtained via \eqn{fsfsL} by transforming the exact solution $s_\L f(s_\L)$ for the effective Gaussian walks.  This confirms that the intuition that the first crossing distribution for Lognormal walks can be mapped to one of Gaussian walks crossing an effective barrier is correct.   

In contrast, for walks with correlated steps (in our case, we chose to smooth a power spectrum $P(k)\propto k^{-1}$ with a Gaussian filter) we must rely on \eqn{eq:fNGeval} or \eqref{fNGrenorm}, both of which approximate $f(s)$ with the probability of crossing upwards (but not necessarily for the first time). However, while the two are formally identical, in the latter all the Hermite terms identically vanish and non-Gaussian effects are treated exactly. Moreover, it is equivalent to approximating $f(s_\L)$ for the Gaussian walks with \eqn{sfs}, which is known to tend to the exact result when the barrier recedes faster than the Gaussian distribution spreads \citep{ms13b}, as is already the case for a linear barrier.
The result is shown by the lower solid curve, while the dashed curve shows \eqn{eq:fNGeval} with all Hermite correction terms set to zero to evaluate the effect of truncating the series. Clearly, both are in good agreement at large $\delta_c^2/s \gg 1$, with \eqn{fNGrenorm} providing a better description at smaller $\delta_c^2/s$, as expected.  We conclude that our analysis is able to provide a good description of the first crossing distribution by non-Gaussian walks even when the non-Gaussianity is large.

%For correlated steps, one still has to use \eqn{sfs}.
%Hence, the first crossing distribution expressed as a function of $s$ is:  
%\be
% % f(s)ds = f(sig2) dsig2
% % f(s) = f(sig2) dsig2/ds
% %                ds/dsig2 = 1+s
% % f(s) = f(sig2)/(1+s)
% (1+s)\,f(s) = 
% \frac{\ln(1+\Delta_c)}{\ln(1+s)}\, 
% \frac{\mathrm{exp}(-B^2_\L/2\sigma^2_\L)}{\sqrt{2\pi}\sigma_\L}.
%\ee

Before moving on, we note that because $b_\L$ increases faster than the width of the Gaussian distribution $\sqrt{s_\L}$, a fraction 
% $\mathrm{e}^{-\ln(1+\delta_c)} = (1+\delta_c)^{-1}$ 
of walks never cross the barrier (as discussed by \citealt{ms13b}).  For sufficiently high barriers, only a tiny fraction of the walks will cross.  However, if the physics of collapse does not depend on whether or not the initial field was Gaussian, then we expect $\delta_c$ to be of order unity in both cases, and the barrier does not become very high.

%For small $\Delta_c$, $\ln(1+\Delta_c)\to\Delta_c$ and this becomes 
%\be
% (1+s)\,f(s) = 
% \frac{\Delta_c}{\ln(1+s)}\, 
% \frac{\mathrm{exp}(-B^2_\L/2\sigma^2_\L)}{\sqrt{2\pi}\sigma_\L}.
%\ee

%The first crossing distribution for these Lognormal walks crossing a barrier of constant height is the same as for Gaussian walks crossing $b_\L = \ln(2)\delta_c + \sigma^2_\L/2\delta_c$, %= \delta_c\ln(2\sqrt{\Sigma})$, upto a factor $\dd\sigma_\L^2/\dd s = (1+s/\delta_c^2)^{-1}$.

As a simple additional example, consider a polynomial transformation $\delta = \delta_\L + \alpha(\delta_\L^2-s_\L)/\sqrt {s_\L} + \beta\delta_\L^3/s_\L$, which returns a variable with zero mean and variance $s = s_\L^2[1 + 2(\alpha^2+3\beta) + 15\beta^2]$. %For appropriate values of $\alpha$ and $\beta$, this transformation is still monotonic, but the distribution of $\delta$ is now skewed.  In this case, a barrier $b$ translates to an effective barrier $b_\L$ for which $b_\L + \alpha b_\L^2/\sigma_\L + \beta b_\L^3/\sigma_\L^2 = b + \alpha\sigma_\L$.  Therefore, the effective barrier that the Gaussian walk $\delta_\L$ must cross will now depend on $\sigma_\L$, even if $b=\delta_c$ is constant.  Since \eqn{sfs} applies to moving barriers too, this poses no problem: all one has to do is to modify the factor $\dd\sigma_\L^2/\dd s$ in \eqn{monotonic} accordingly.
Tuning the values of $\alpha$ and $\beta$, one can use this transformation to reproduce the skewness and kurtosis of any distribution, and in particular to mimic the effect of primordial non-Gaussianity of a given type (see for instance \citealt{MVJ00}, or more recently \citealt{mp11}. Notice, however, that the correlations between scales in the resulting model are inherited from the underlying Gaussian walks, and the full three-point function at different scales may not, and in general will not, be reproduced correctly).  
%Because the approximation described in this paper only depends on same-scale quantities, this does not matter in the large mass regime. It will however matter for small masses, and whenever corrections to \eqn{fmeanv} are needed. Figure ... shows the agreement between the exact toy models presented here and the different approximations described in the previous Sections.

\subsubsection{Further generalization}
As we noted above, a similar rescaling (with no additional corrections) occurs whenever the conditional distribution $p(v|b)$ remains Gaussian, even though $p(b)$ is not. This is the case of most phenomenological models used in Cosmology, with at most small perturbations around a Gaussian distribution.
In order to construct a model for which the second line of \eqn{fNGrenorm} becomes large, one should consider for instance a transformation of the kind
\begin{equation}
  \label{weirdv}
  \delta=\delta_\L + \delta_c
  \int_0^{s_\L}\dd S_\L\left(v_\L^2-\avg{v_\L^2}\right) \,,
\end{equation}
which once differentiated reads
\begin{equation}
  v = \frac{\dd s_\L}{\dd s} 
  \left[v_\L + \delta_c\left(v_\L^2-\avg{v_\L^2}\right)\right].
\end{equation}
In such a transformation, $v_\L$ must always appear in an integral so that no additional stochastic variables are introduced, and it must appear at least quadratically so that the relation with $v$ is non-linear.  We are not aware of any phenomenological motivation for tranformations such as \eqn{weirdv}, so we have not pursued this further.

%%%%%%%%%%%%%%%%%%%%%%%%%%%%%%%%%%%%%%%%%%%%%%%%%%%%%%%%%%%%%%
\section{Applications}\label{sec:other}
%%%%%%%%%%%%%%%%%%%%%%%%%%%%%%%%%%%%%%%%%%%%%%%%%%%%%%%%%%%%%%
The analysis above was motivated by the possibility that the initial density fluctuation field was non-Gaussian.  We discuss this first, and then show that we can use our results to address a number of other interesting problems as well.

%%%%%%%%%%%%%%%%%%%%%%%%%%%%%%%%%%%%%%%%%%%%%%%%%%%%%%%%%%%%%%
\subsection{Halo abundances in models with primordial non-Gaussianity}\label{sec:png}
%%%%%%%%%%%%%%%%%%%%%%%%%%%%%%%%%%%%%%%%%%%%%%%%%%%%%%%%%%%%%%
Primordial non-Gaussianity is expected to be small \citep{planckNG}.  This means that we can work with \eqn{eq:fNGeval} rather than \eqn{fNGrenorm}.  In addition, for both local and equilateral models of non-Gaussianity, $p(B)$ is only known from its moments, via equations~(\ref{pBexpW}) and~(\ref{W}), which depend on $B$ and $\avg{\Delta^3}_c$.  As discussed by \citet{damico11}, for values of $\avg{\Delta^3}_c\sim 0.01$ and $B\sim 10$ (corresponding to the most massive clusters of galaxies) the three terms listed in Eq.~\eqref{W} are $\mathcal{O}(100)$, $\mathcal{O}(10)$ and $\mathcal{O}(1)$ respectively, while neglected terms start with $\mathcal{O}(10^{-1})$.  These values are obtained for primordial non-Gaussianity with $f_{\mathrm{NL}}\sim 100$, which is now excluded by the Planck mission \citep{planckNG}.  However, since even larger values of $B$ can be attained at higher redshifts, or by the study of different objects like the reionisation pattern of cosmic structures \citep{Joudaki11,D'Aloisio2013,lidz13}, the previous discussion about how to truncate $W(B)$, besides having its own theoretical interest, was not unnecessary.  In any case, this shows that the relevant large mass limit of $f(s)$ is given by \eqn{fsW}.

Most previous work on non-Gaussian excursion sets has considered a barrier of constant height, for which $-B' = B/2s$. This is for instance the case of \citet{MVJ00} and \citet{lmsv08}, who also explicitly assume that $f(s)=2f_\PS(s)$ (see however the discussion on the fudge factor by \citeauthor{MVJ00}). This assumption, together with the choice of a Top-Hat filter like the one they use, does not appear justified from the point of view of excursion sets. However, their main concern was reproducing the results of N-body simulations, rather than excursion sets, and multiplying by 2 was going in the correct direction. Also, this error disappears when considering the non-Gaussian to Gaussian ratio, as they did, with the aim of computing the correction that should multiply the result Gaussian simulations. 

The correspondence between our expression in the large mass limit and theirs is helped by noting that it is conventional to define $\sigma S_3\equiv \avg{\!\Delta^3\!}_c$, so our $\avg{\!\Delta^3\!}_c' \equiv \dd(\sigma S_3)/\dd s$.  Since \eqn{fPS} with an extra factor of 2 is the full story, they are in effect missing the $X$ dependent corrections which matter at smaller masses. 

Our large mass limit differs slightly from that of \citeauthor{lmsv08} only because they keep the Edgeworth expansion, while we have been careful about how we write the large mass limit of \eqn{pBexpW}. In this we followed \citet{damico11} who pointed out that perturbative non-Gaussian corrections blow up at small $s$, and need to be resummed in an exponential, whose argument corresponds to \eqn{W} in this regime.  The same approach is followed by \citet{LoVerde11}.  

If $\sigma S_3$ is only weakly scale-dependent, so the $\avg{\!\Delta^3\!}_c'$ term can be dropped, then $f(s)$ satisfies \eqn{fNGratio}.  \citet{mp11} used this to argue that the large scale limit of the non-Gaussian mass function from correlated random walks is always one half of the one obtained without filter-induced correlations, finding very good agreement with Monte-Carlo simulations.  They also pointed out that this factorisation justifies the common practice of obtaining the full non-Gaussian mass function as the product of the fit from Gaussian simulations times an analytically predicted non-Gaussian correction. Our results confirm this intution, and at the same time highlight the conditions under which it holds true.  E.g., our analysis shows that writing $f(s)$ in this way is not appropriate at lower masses, nor will it be accurate if $\sigma S_3$ is scale-dependent.

From Figure~1 of \citet{damico11}, in whose notation $\avg{\!\Delta^n\!}_c=\epsilon_{n-2}$, one can infer that for local non-Gaussianity $\avg{\!\Delta^3\!}_c\simeq (2 \times 10^{-4}) f_{\mathrm{NL}}$ and $\avg{\!\Delta^4\!}_c\simeq 10^{-7} f_{\mathrm{NL}}^2+ (2 \times 10^{-8}) g_{\mathrm{NL}}$, both quantities being nearly constant.  For non-Gaussianity of the local type the ratio gives
\begin{equation}
  \frac{f_{\mathrm{NG}}(s)}{f_{\mathrm{Gauss}}(s)} \simeq
  \exp\bigg[ \frac{f_{\mathrm{NL}}}{30}\frac{B^3}{10^3}
  + \frac{(g_{\mathrm{NL}}-f_{\mathrm{NL}}^2)}{1.2\times10^5}\frac{B^4}{10^4}
  +\dots\bigg].
\label{ratiofnl}
\end{equation}
Interestingly, this is the same as the exponential part of equation (34) of \citet{damico11}, which however was derived for uncorrelated steps and thus has a factor of 2 in what they call $f_{\mathrm{PS}}$. Even more interestingly, this corresponds to the first line of their equation (44) assuming that their factor $(1-\tilde\kappa+\dots)$ due to filter corrections -- obtained following \citet{mr10c} -- resums exactly to 1/2 (while setting their $D_B=0$ one gets $1-\tilde\kappa=1-\kappa=.54$). Differences start emerging only at the order $\epsilon_1\nu$: including the first terms neglected in \eqn{ratiofnl} would give in their language $(2-c_1)/2$ instead of $(4-c_1)/4$ as the coefficient for this term.
%, and that the term in braces should resum to 1 at very large masses. However, the comparison is partially blurred by the fact that some of theirs corrections come from walks with earlier crossings, which we do not include here. 
Since these corrections are anyway quite small (much less than one percent on the scales of interest) this substantially confirms their result, in a much simpler and more intuitive way.

Moving barriers and weakly non-Gaussian fields were first considered by \citet{ls09}, but only for a sharp-k filter.  They found that, for moving barriers also, the large mass limit is just the Gaussian result times the non-Gaussian correction to the pdf, provided $\dd (\sigma S_3)/\dd s$ is small.  Our more general analysis confirms this is true for other filters also, although the Gaussian result itself depends on the smoothing filter.  

A self-consistent treatment of excursion sets with correlated steps was laid out by \citet{mr10c}, who used a path integral formalism to compute the first crossing rate for barriers of constant height. Unfortunately, their choice to expand around the uncorrelated Gaussian solution makes the calculations very involved. For their choice of filter (Top-Hat) and matter power spectrum ($\Lambda$CDM), the first non-Markovian correction brings their Gaussian result within 10\% of the correct answer, but the reliability of the non-Gaussian corrections becomes problematic for large masses (while we instead recover the exact large mass limit already at leading order). Moreover, in their framework a change in the correlation function (like using a different filter or power spectrum) would require to redo the calculations.
The same is true for other works following the same approach like \citet{ca11} (who were only able to consider linear barriers with small slope) and \citet{damico11} (who did not consider moving barriers, but focussed on the safer non-Gaussian to Gaussian ratio, and assessed the range of validity of their results). 

In conclusion, while there are sometimes small differences in detail, all analyses suggest that the only regime where one might hope to detect primordial non-Gaussianity is at large masses.

%%%%%%%%%%%%%%%%%%%%%%%%%%%%%%%%%%%%%%%%%%%%%%%%%%%%%%%%%%%%%%
\subsection{Stochastic barrier models}
\label{sec:stocNG} 
%%%%%%%%%%%%%%%%%%%%%%%%%%%%%%%%%%%%%%%%%%%%%%%%%%%%%%%%%%%%%%

It is well known that the physics of halo formation depends on more than the initial overdensity field -- the shear field associated with a proto-halo patch also plays an important role \citep{bm96,smt01}.  Following \cite{st02}, models of this process typically search for the largest smoothing scale on which 
\begin{equation}
 \label{dandq}
 \delta_\L \ge \delta_c\,(1 + q/q_c) ,
\end{equation}
where $q$ is an independent stochastic variable (or variables) that is typically non-Gaussian.  Such models are usually described in terms of multiple-dimensional random walks crossing a barrier, and have a considerably richer structure than the one-dimensional walks we have considered in the paper \citep{cs13}.  Upon defining 
\begin{equation}
  \delta\equiv \delta_\L - \beta q \, , 
\end{equation}
with $\beta=\delta_c/q_c$, the problem reduces to finding the first crossing distribution of a constant barrier $\delta_c$ by the non-Gaussian variate $\delta$, whose correlation structure is inherited from those of $\delta_\L$ and $q$.  

For instance, \cite{scs13} study a model in which $q^2\equiv\sum_{i=1}^5\delta_i^2/5$ with $\avg{\delta_i^2}=\avg{\delta^2_\L}\equiv s_\L$ and $\avg{\!\delta_i\delta_j\!}=0$. That is, they assume that $q^2$ is independent of $\delta$ and is drawn from a Chi-squared distribution with 5 degrees of freedom, each of which is correlated like $\delta$. 
In \cite{ms14} we show that, because $q$ is built from Gaussian walks, this is a case for which also truncating \eqn{eq:fNGeval}, neglecting the Hermite polynomials, actually leads to the exact result (for $\delta$ upcrossing a barrier).  Moreover, it has $f_\PS(s)=-B'p(B)$, which means that the truncated \eqn{eq:fNGeval} is actually just \eqn{sfs}, with $p_\G(B)$ replaced by $p(B)$.  This is a remarkably simple result for these physically motivated non-Gaussian walks.

%%%%%%%%%%%%%%%%%%%%%%%%%%%%%%%%%%%%%%%%%%%%%%%%%%%%%%%%%%%%%%
\subsection{Halo abundances from the nonlinear field}
\label{sec:nlPDF}
%%%%%%%%%%%%%%%%%%%%%%%%%%%%%%%%%%%%%%%%%%%%%%%%%%%%%%%%%%%%%%

The excursion set approach was formulated to predict the abundance of nonlinear objects from the initial fluctuation field.  However, because \eqn{fNGrenorm} is valid even if $p(b)$ is highly non-Gaussian, we can use it to predict halo abundances from the late time field as well.  The problem is particularly simple because halos are often identified in the nonlinear field by finding a spherical or triaxial patch that is a fixed multiple of the background density, independent of halo mass \citep{dts13}.  In effect, this means our excursion set approach, applied to the nonlinear non-Gaussian field with a constant barrier, is an analytic model of the numerical halo finding algorithm.

This has an important consequence for studies of the halo distribution that seek to approximate the smoothed halo field as a Taylor series in quantities derived from the underlying matter distribution.  If the Taylor series is in the matter overdensity only, then the halos are said to be locally biased with respect to the mass.  Our analysis shows that the bias {\em must} be nonlocal since the mass overdensity is not the only quantity which matters:  at the very least, the first derivative of the matter field with respect to smoothing scale plays an important role in determining halo abundances, and this is expected to make the halo-mass bias $k$-dependent \citep{ms12}.

That said, the critical nonlinear overdensity is of order $100\times$ the background.  This is substantially (at least $10\times$) larger than the rms value of the field when smoothed on the typical halo scale, so it may be that the additional terms which come from constraining the slope are irrelevant.  Since in this limit, our \eqn{fNGrenorm} reduces to \eqn{fPS}, the halo mass function is very simply related to the probability distribution function of the nonlinearly evolved field.  We are in the process of exploring this nonlinear excursion set approach further.

To do this quantitatively, we must face the question of how precisely to implement the excursion set approach in the nonlinear field.  The most naive application would mean that we are interested in the smallest $s$ at which $\delta\ge \delta_c$.  To see why this will be problematic, consider the limit in which the density field is made of halos which are negligibly small compared to the spaces between them.  Then, if we set $\delta_c\gg 1$, only those cells which are precisely centered on a halo will have $\delta\ge\delta_c$; the vast majority of randomly placed cells will have $\delta\ll \delta_c$.  Since the excursion set approach equates the fraction of such cells to the mass fraction in halos, it will predict that only a small fraction of the total mass is bound up in halos.  E.g., for the Lognormal distribution described above, the excursion set approach with $p(\delta,s)$ will have a first crossing distribution which looks just like that for Gaussian walks crossing a linearly increasing barrier $b_\L = \ln(1+\delta_c) + s_\L/2$.  For uncorrelated steps it is well-known that only a fraction $\mathrm{e}^{-\ln(1+\delta_c)} = (1+\delta_c)^{-1}$ of the walks will cross such a barrier (similar result applies to correlated steps, as shown by \citealt{ms13b}), leading to the incorrect conclusion that only a small fraction of the mass is bound-up in halos.  

Indeed, in practice, tests of the excursion set approach correspond to studying the statistics of walks that are centered on randomly chosen particles of the distribution (rather than randomly chosen positions).  A crude way to account for this is to mass-weight the walks when estimating the first crossing distribution (e.g. Sheth 1998).  This would be exact for the point-cluster model (Abbas \& Sheth 2007).  That is to say, the non-Gaussian distribution which should be inserted into the excursion set formula is not $p(\delta,s)$ itself, but $(1+\delta)\,p(\delta,s)$.  
The result of mass-weighting the distribution is particularly easy to see for the Lognormal, since $1+\delta = \mathrm{e}^{\ln(1+\delta)}$ so 
$(1+\delta_c)\,p(\delta_c,s)$ looks like 
$\exp(-\bar b_\L^2/2s_\L)/\sqrt{2\pi s_\L}$ 
where $\bar b_\L\equiv \ln(1+\delta_c) - s_\L/2$.  Thus, mass weighting the walks is like studying Gaussian walks crossing a linearly decreasing (rather than increasing) barrier, so all walks are guaranteed cross. 

% ln(rho) - [ln(rho) + sig2/2]^2/2/sig2
% = [2 ln(rho) sig2  - ln(rho)^2 - sig2^2/4 - 2 ln(rho) sig2/2]/2/sig2
% = [2 ln(rho) sig2/2  - ln(rho)^2 - sig2^2/4]/2/sig2
% = -[ln(rho) - sig2/2]^2/2/sig2

The analysis above has a nice connection to previous work, and so allows one to address a related question, having to do with the self-consistency of the approach.  Namely, suppose we estimate halo abundances not from the initial field (as is usually done), but from a weakly evolved one.  How does the prediction compare with that based on the initial field (the usual estimate)?  If the approach is self-consistent, these two estimates should agree.

Let $p(M|V_\mathrm{E})$ denote the probability that a cell of volume $V_\mathrm{E}$ placed randomly in the evolved distribution contains mass $M$.  This distribution has mean $\bar{M}\equiv\bar\rho V_\mathrm{E}$, where $\bar\rho$ is the comoving background density, so the Eulerian density is $1+\delta_\mathrm{E}\equiv M/\bar{M}$.  Local models of the evolution from the initial Lagrangian density $\delta_\L$ to the Eulerian one assume that $1+\delta_\mathrm{E}$ is a deterministic invertible function of $\delta_\L$.  In perturbation theory, this means that
\begin{equation}
  \int_M^\infty \dd m\,p(m|V_\mathrm{E})\,(m/\bar{m})
  = \int_{\delta_\L(M,V_\mathrm{E})}^\infty \dd\delta\,
  p(\delta|s(M))
 \label{matchPT}
\end{equation}
\citep{bcgs02,ls08}. (If halos were made of discrete particles, then this mass weighting is similar to only counting cells which are centered on particles of the distribution.)  Halos correspond to large $M/\bar\rho V_\mathrm{E}$ for which $\delta_\L(M,V_\mathrm{E})\to \delta_c(M)$.  Since the right hand side here is the same as the right hand side of \eqn{fPS}, the extra factor of $M/\bar M$ on the left hand side here shows that it is the mass-weighted Eulerian distribution which is related to the halo mass function.  But mass-weighting the walks was is exactly what we argued was necessary to make sense of the excursion set predictions, so it is reassuring that this is, in fact, what equation~\eqref{matchPT} does.  This demonstrates self-consistency at least at the higher masses, where \eqn{fPS} is the appropriate limit of the two (i.e., the Eulerian and Lagrangian) predictions.

%%%%%%%%%%%%%%%%%%%%%%%%%%%%%%%%%%%%%%%%%%%%%%%%%%%%%%%%%%%%%%%%%%

\section{Discussion}

%%%%%%%%%%%%%%%%%%%%%%%%%%%%%%%%%%%%%%%%%%%%%%%%%%%%%%%%%%%%%%%%%%

We derived an intuitively simple formal expansion for the first crossing distribution of random walks with correlated steps, in which walks are ordered by the number of times they cross the barrier from below (equation \ref{eq:corrections}). The nature of the correlations between the steps is determined by the statistics of the field (i.e. Gaussian or non-Gaussian) when it is smoothed, which itself depend on the form of the smoothing filter.  The leading order term of this expansion, \eqn{eq:fNGeval}, is particularly simple.  It only requires that when walks cross the barrier, they do so crossing upwards.  Therefore, it requires knowledge of only the joint distribution of the walk height and its first derivative:  in appropriately scaled units, these turn out to be independent of one another (equation~\ref{pgauss}), making the analysis particularly simple.  (Figure~\ref{fupnc} and associated discussion illustrate a simple approximation for treating the multiple upcrossings problem.)

Previous work has shown that, for Gaussian initial conditions, the simplest approximation (i.e. neglecting all the other terms associated with walks with multiple zig-zags) leads to \eqn{sfs}, which works well for all filters of current interest, and for all barriers which are monotonic functions of smoothing scale.  Our \eqn{eq:fNGeval} is a straightforward generalization of \eqn{sfs} to non-Gaussian fields:  again, only the bivariate distribution of height and slope is required.  In the large mass regime, our formula reduces to the even simpler form of \eqn{fPS}, which depends on the distribution of the walk heights alone.  Despite the fact that perturbative non-Gaussian corrections individually blow up in this regime, this result is completely non-perturbative and exact:  it simply reflects the fact that those walks that reach the barrier in very few steps are very unlikely to have crossed it multiple times, because of the correlations.

Our \eqn{eq:fNGeval} involves an infinite series, truncation of which is reasonable if the non-Gaussianity is weak.  For this reason, we provided another expansion, \eqn{fNGrenorm}, which is more efficient in the case of large non-Gaussianity.  When the non-Gaussian field is obtained by making a deterministic transformation of a Gaussian field, then the difference between these two expansions is particularly easy to see.  In such cases, the first crossing problem becomes particularly simple:  the non-Gaussian problem can be mapped to one of Gaussian walks crossing an effective barrier whose shape is related to the original one (equation~\ref{fsfsL}).  We illustrated the argument using a few simple examples, of which the Lognormal is particularly instructive (Figure~\ref{lnWalks} and associated discussion).  

Equation~(\ref{eq:fNGeval}) is useful for excursion set models which assume that the initial fluctuation field was non-Gaussian; indeed, this was the original motivation for this study.  We discussed compared our results to previous work on this topic in Section~\ref{sec:png}.  However, our analysis is unique in that we do {\em not} assume that the non-Gaussianity is weak.  Therefore, our equations~(\ref{eq:fNGeval}) and~(\ref{fNGrenorm}) can be used to predict the abundance of nonlinear objects in two other cases of interest.  Section~\ref{sec:stocNG} discussed the case in which halo formation depends on quantities other than the initial overdensity.  E.g., in the triaxial collapse model, the relevant quantity is obtained by convolving $\delta$ with a non-Gaussian variate (equation~\ref{dandq}), so that the problem to be solved reduces to one of non-Gaussian walks crossing a simple barrier.  This is a rather different view of what is usually regarded as a problem involving walks in multiple-dimensions, so we expect this to be particularly useful for future models of sheets and filaments in the cosmic web.  

The other application is to predict halo abundances from the nonlinear (i.e. late time) rather than the initial fluctuation field.  Section~\ref{sec:nlPDF} argued that this means that halo bias must be nonlocal in principle, although local bias may be a good approximation in practice.  We also argued that our formulation demonstrates self-consistency of the approach, in the sense that applying it to the initial or the late-time field (i.e., the Lagrangian or Eulerian fields) yields the same estimate of halo abundances, at least at the higher masses where \eqn{fPS} is the appropriate limit.  However, this self-consistency requires that one mass weight the Eulerian space walks to which the excursion set argument is applied -- in the Lagrangian formulation, all walks have the same weight (equation~\ref{matchPT} and related discussion). It will be interesting to explore if this self-consistency survives in modified gravity models, where, because the linear theory growth factor becomes $k$-dependent, even just the linearly evolved field is rather different from the initial one.

Finally, we noted in the Introduction that the fundamental excursion set ansatz \eqn{ansatz}, the primary motivation for interest in barrier crossing problems in cosmology, is known to be incorrect.  This is because the excursion set prediction for the mass of the halo in which a particle will be at some later time is only accurate around special positions in the initial field \cite[those around which collapse occurs;][]{smt01} whereas the fundamental ansatz assumes that it is accurate around all positions.  In principle, this undermines interest in the problems we have addressed in this paper.  However, as \cite{ms12} have argued, the result of including only the relevant subset of walks over which to average boils down to weighting some walks more than others.  E.g., the excursion set peaks patch model for halo formation \citep{bm96, psd13} corresponds to applying an additional weight which depends on the slope of the walk when it crosses the barrier (what we called $v$) \citep{ms12, ps12}.  Since no change, other than this additional weighting term, must be made to our upcrossing formalism, we expect our analysis of first crossing distributions for non-Gaussian to be useful, and to motivate further study of precisely what is special about those positions in space around which collapse occurs.

%%%%%%%%%%%%%%%%%%%%%%%%%%%%%%%%%%%%%%%%%%%%%%%%%%%%%%%%%%%%%%%%%%

\section*{Acknowledgments}

%%%%%%%%%%%%%%%%%%%%%%%%%%%%%%%%%%%%%%%%%%%%%%%%%%%%%%%%%%%%%%%%%%

MM is supported by the ESA Belgian Federal PRODEX Grant No.~4000103071 and the Wallonia-Brussels Federation grant ARC No.~11/15-040.  He is grateful to the University of Pennsylvania for its hospitality in December 2013.  RKS was supported in part by NSF-AST 0908241.  He is grateful to the Perimeter Institute for its hospitality in September 2013, G. Dollinar and D. Bujatti for their hospitality in Trieste during October 2013, and the Institut Henri Poincare for its hospitality in November 2013.

\appendix

%%%%%%%%%%%%%%%%%%%%%%%%%%%%%%%%%%%%%%%%%%%%%%%%%%%%%%%%%%%%%%%%%%
%%%%%%%%%%%%%%%%%%%%%%%%%%%%%%%%%%%%%%%%%%%%%%%%%%%%%%%%%%%%%%%%%%

\section{Multiple upcrossings}

\label{sec:corrections}

%%%%%%%%%%%%%%%%%%%%%%%%%%%%%%%%%%%%%%%%%%%%%%%%%%%%%%%%%%%%%%%%%%

In this appendix we discuss the derivation of \eqn{fmeanv} in a more formal way, and calculate the corrections one must account for when $s$ becomes too large, or simply if one wants to quantify the errors introduced by the approximations we made.

The result above can be derived in a more rigorous way within a path integral formulation of the excursion set theory, where one considers an ensemble of walks of $N$ steps with infinitesimal increment in variance $\Delta s=s/N$.  The first crossing rate is by definition the fraction of walks that crossed for the first time at the last step, over the width of the step. Calling $p(\delta_1,\dots,\delta_N)$ the joint probability of a walk, this is 
\begin{equation}
  f(s)=\frac{1}{\Delta s} \int_{-\infty}^{b_1}\!\!\!\!\dd\delta_1\dots\!
  \int_{-\infty}^{b_{N-1}}\!\!\!\!\!\!\!\!\!\!\!\dd\delta_{N-1}\!
  \int_{b_N}^{\infty}\!\!\!\dd\delta_N \,
  p(\delta_1,\dots,\delta_N)
\end{equation}
where $b_i\equiv b(s_i)$ is the value of the barrier at the scale $s_i=i\Delta s$ corresponding to the $i$-th step.

This expression can be written as the difference of two path integrals: a first one including all possible values of $\delta_1,\dots,\delta_{N-2}$, and a second one removing walks with at least one $\delta_i>b_i$. The former is marginalized over $\delta_1,\dots,\delta_{N-2}$, and thus is just the probability of having $\delta_{N-1}<b_{N-1}$ and $\delta_N>b_N=b$, normalized to $\Delta s$. This is equal to 
\begin{align}
   \frac{1}{\Delta s}
  \int_{b'}^{+\infty}\!\!\!\dd v
  \int_b^{b+(v-b')\Delta s}\!\!\!\!\!\dd\delta_N \,p(\delta_N,v)\,,
\end{align}
where $v\equiv(\delta_N-\delta_{N-1})/\Delta s$ and $b'\equiv (b_N-b_{N-1})/\Delta s$.
For correlated steps, $\avg{\!(\delta_N-\delta_{N-1})^2\!}\propto\Delta s^2 $: all correlators of $v$ and $\delta_N$ tend to constants, and so does $p(\delta_N,v)$. The $\Delta s\to0$ limit of this term thus returns the r.h.s.~of \eqn{fmeanv}.

The domain of the second path integral, that corrects the error introduced by the marginalizations, is for the first $N-2$ steps $\prod_{i=1}^{N-2}\big(\int_{-\infty}^{\infty}\dd\delta_i - \int_{b_i}^{\infty}\dd\delta_i\big)- \prod_{i=1}^{N-2}\int_{-\infty}^{+\infty}\dd\delta_i$.
Up to an overall minus sign, this is equal to
\begin{equation}
  \sum_{i=1}^{N-2}\int_{-\infty}^{b_1}\!\!\!\dd\delta_1\dots
  \int_{-\infty}^{b_{i-1}}\!\!\!\!\!\!\dd\delta_{i-1}
  \int_{b_i}^{\infty}\!\!\!\dd\delta_i
  \int_{-\infty}^{\infty}\!\!\!\!\dd\delta_{i+1}\dots\dd\delta_{N-2}\,,
\end{equation}
where the $i$-th term of the sum removes all the trajectories that had crossed for the first time at $s_i<s_{N-1}$. 
One can then iterate the procedure, marginalizing the first $i-2$ variables of each term, at the cost of introducing a new term with a sum up to $i-2$ to correct the error, and so on. This yields an alternating series, in which the $k$-th term with $k$ nested sums corrects for the trajectories with $k$ crossings miscounted in the previous terms.

If we repeat the same considerations for each of the $k$ earlier crossings, introducing the velocities $v_1,\dots,v_k$ in addition to $v_{k+1}\equiv v$, we obtain the continuum limit by replacing the $k$ sums $\sum\Delta s$ with nested integrals over the crossing scales $s_1<\dots<s_k<s\equiv s_{k+1}$. Finally, we divide by $k!$ and drop the constraint on the ordering of $s_1,\dots,s_k$. This gives 
\begin{align}
  f(s) &= \int_{b'}^{+\infty} \!\!\mathrm{d}v \, (v-b') \, p (b,v)
  \notag \\
  &+\sum_{k=1}^{\infty}\frac{(-1)^k}{k!}\!
  \int_0^s\!\!\dd s_1\dots\int_0^s\!\!\dd s_k \,
  f^{(k)}(s_1,\dots,s_k,s)\,,
\label{fsall}
\end{align}
where [calling $s\equiv s_{k+1}$ and $b(s) \equiv b_{k+1}$]
\begin{equation}
  \label{fk}
  f^{(k)} = \int_{b'}^{\infty} \!\!\dd v  \dots 
  \int_{b'_1}^{\infty} \!\!\dd v_1
  \prod_{j=1}^{k+1}(v_j-b_j') \, p(\{b_i,v_i\})\,.
\end{equation}
\citet{ms12} keep only the leading term of the full expression for $f(s)$, while stopping the expansion at $k=1$ yields \eqn{eq:corrections}.

\begin{figure}
 \centering
 \includegraphics[width=0.9\hsize]{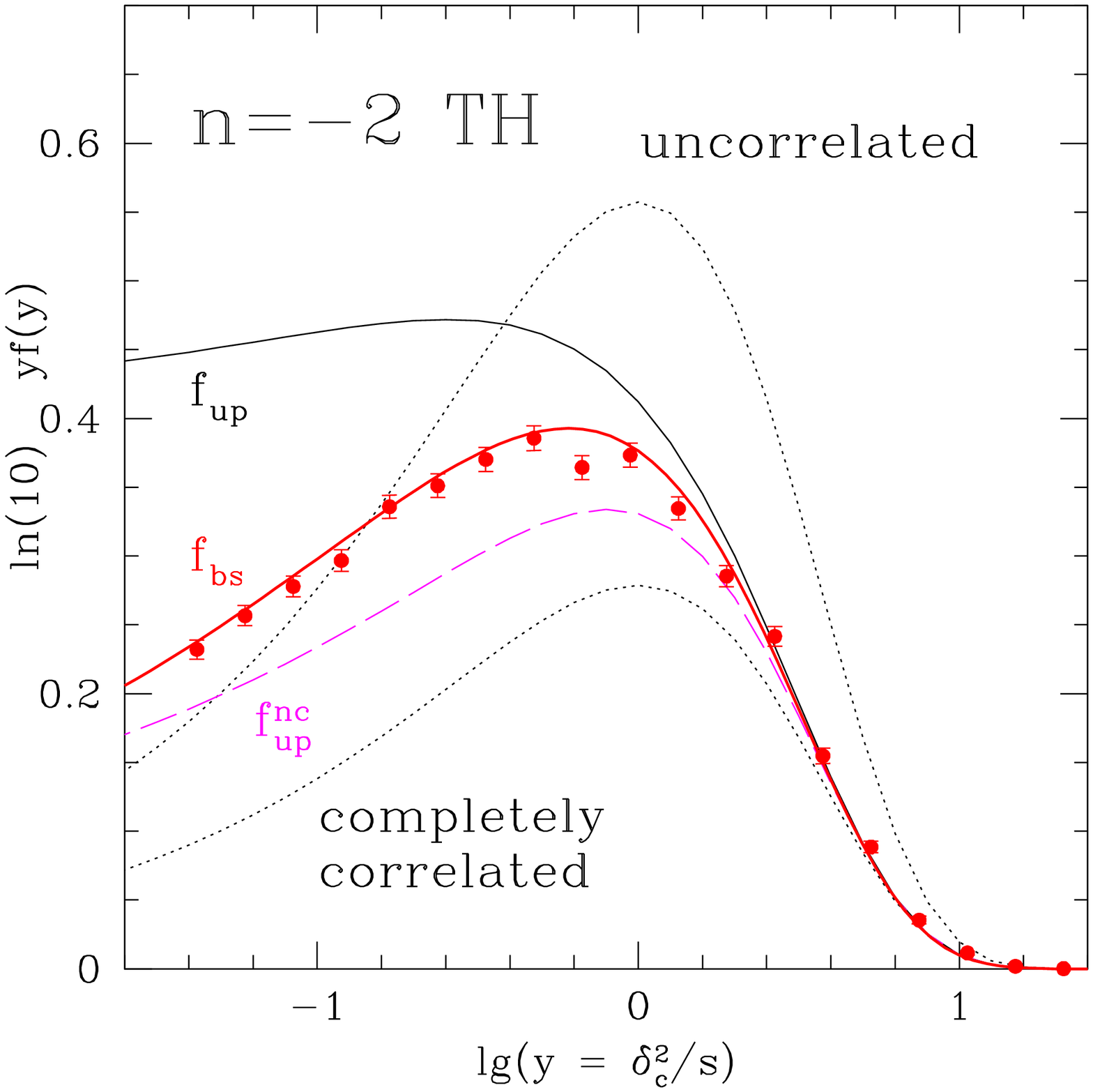}
 \caption{\label{fupnc} First crossing distribution of a barrier of constant height $\delta_c$ by Gaussian walks having steps that are correlated because of TopHat smoothing of a power spectrum with $P(k)\propto k^{-2}$ (symbols with error bars).  Dotted curves show $f_\PS$ of \eqn{fPS} and $2f_\PS$; thin solid curve shows $f_{\mathrm up}$ of \eqn{sfs}; dashed curve shows the no-correlation approximation of \eqn{simplestcase}; and thick solid curve shows the back-substitution solution of \cite{ms13b}.}
\end{figure}

To see what these expressions imply, suppose that $p(\{b_i,v_i\}) = \prod_i p(b_i,v_i)$; this keeps the correlation between the walk height and its slope on each scale, but assumes that these are uncorrelated with the height and slope on any other scale.  This makes the integrals over $s_k$ separable, so the result is the product of $k$ terms:
\begin{equation}
  f(s) = f_{\rm up}(s)
  \bigg[1 + \sum_{k\ge 1} \frac{(-1)^k}{k!}\,
                     \prod_{j=1}^k\,f_{\rm up}(<s;b_j)\bigg],  
\end{equation}
where $f_{\rm up}(<s) = \int_0^s \dd s\,f_{\rm up}(s)$, and $f_\mathrm{up}$ is the leading term in \eqn{fsall} (equation~\ref{fmeanv} in the main text).  If we further assume that the barrier was constant, then each of the terms in the product is the same, making
\begin{equation}
 \label{simplestcase}
 f(s) = f_{\rm up}(s) \,\exp[-f_{\rm up}(<s)]\,.
\end{equation}
This final expression is the same as equation~(A3) of \citet{bcek91}.

Since $f_{\rm up}(<s)$ increases as $s$ increases -- and even exceeds unity at large enough $s$ -- the result of including the extra terms is to damp $f_{\rm up}(s)$ at large $s$.  E.g., the expression above indicates there will be an approximately 15\% correction downwards at $s\simeq\delta_c^2$.  This turns out to be slightly larger than the actual correction (see Figure~\ref{fupnc}) because in fact, $p(\{b_i,v_i\})\ne \prod_i p(b_i,v_i)$.  Although including the additional corrections which come from correlations between scales complicate the analysis \cite[see][]{ms13b}, we believe the algebra above illustrates nicely how the inclusion of multiple upcrossings will impact the result as $s$ increases.

%%%%%%%%%%%%%%%%%%%%%%%%%%%%%%%%%%%%%%%%%%%%%%%%%%%%%%%%%%%%%%%%%%
\section{The non-Gaussian PDF}
\label{sec:pdf}
%%%%%%%%%%%%%%%%%%%%%%%%%%%%%%%%%%%%%%%%%%%%%%%%%%%%%%%%%%%%%%%%%%

The non-Gaussian probability distribution can be obtained applying a differential operator to its Gaussian counterpart, as
\begin{equation}
  p(B;s)= \mathrm{e}^{\D}\,\frac{\mathrm{e}^{-B^2/2}}{\sqrt{2\pi}}
\end{equation}
where $\D=\sum_{i=3}^{\infty}[\avg{\!\Delta^i\!}_c/i!](-\pd/\pd B)^i$. Expanding the exponential gives
\begin{align}
  \mathrm{e}^{\D} =
  &\,1 +\sum_{i=3}^{\infty}\frac{\avg{\!\Delta^i\!}_c}{i!}
  \,(-\partial_B)^i 
           \notag\\
  & +\frac{1}{2!}\sum_{i,j=3}^{\infty}
    \frac{\avg{\!\Delta^i\!}_c\avg{\!\Delta^j\!}_c}{i!\,j!}
  \,(-\partial_B)^{i+j} 
  +\dots\,.
\end{align}

The probability distribution can then be written in terms of the Hermite polynomials $H_n(B)\equiv {\rm e}^{B^2/2}(-\partial_B)^{n}{\rm e}^{-B^2/2}$, as
\begin{align}
  p(B;s) &= \,\frac{\mathrm{e}^{-B^2/2}}{\sqrt{2\pi}}
   \bigg[1 +\sum_{i=3}^{\infty}\frac{\avg{\!\Delta^i\!}_c}{i!}\,H_i(B) 
      \notag\\
  & +\sum_{i,j=3}^{\infty}
    \frac{\avg{\!\Delta^i\!}_c\avg{\!\Delta^j\!}_c}{2!\,i!\,j!} 
    \,H_{i+j}(B)
      \notag \\
  & +\sum_{i,j,k=3}^{\infty} \!\!
  \frac{\avg{\!\Delta^i\!}_c\avg{\!\Delta^j\!}_c\avg{\!\Delta^k\!}_c}{%
      3!\,i!\,j!\,k!} \,H_{i+j+k}(B) 
    +\dots\bigg]
\label{eq:pdf}
\end{align}
which is the Gram-Charlier expansion usually referred to in the literature.

Although formally correct, this expression is not convenient to deal with very large masses. In this regime, $B$ can be so large that one might also have $\avg{\!\Delta^3\!}_c\,B^3\gg1$ (see \citet{damico11} for a detailed discussion), and in order to make reliable predictions one cannot truncate the series but needs to sum an infinite number of terms. 
In order to avoid doing this, it is convenient to resum the series above into an exponential and write
\begin{equation}
  p(B;s)=\frac{\mathrm{e}^{W(B;s)}}{\sqrt{2\pi}}\,,
\end{equation}
where the function $W$ is 
\begin{align}
  W(B;s) &= -\frac{B^2}{2} 
  +\sum_{i=3}^{\infty}\frac{\avg{\!\Delta^i\!}_c}{i!}\,H_i(B) 
      \notag\\
  & +\frac{1}{2!}\sum_{i,j=3}^{\infty}
    \frac{\avg{\!\Delta^i\!}_c\avg{\!\Delta^j\!}_c}{i!\,j!}
    \,h_{ij}(B) 
      \notag \\
  & +\frac{1}{3!}\sum_{i,j,k=3}^{\infty} \!
    \frac{\avg{\!\Delta^i\!}_c\avg{\!\Delta^j\!}_c
    \avg{\!\Delta^k\!}_c}{i!\,j!\,k!} \,h_{ijk}(B) 
   +\dots
\label{eq:W}
\end{align}
and where we have defined the modified polynomials
\begin{align}
  h_{ij} &\equiv H_{i+j}- H_iH_j\,, \\
  h_{ijk} &\equiv H_{i+j+k}- H_iH_jH_k - (H_i \,h_{jk}+\mathrm{perms.})
  \notag\\
  & = H_{i+j+k} + 2H_iH_jH_k - (H_i H_{j+k}+\mathrm{perms.})\,, \\
  h_{ijkl} &\equiv H_{i+j+k+l}- H_iH_jH_kH_l \notag\\
  &\phantom{\equiv}\; - (H_i \,h_{jkl}+\mathrm{perms.})
            -(h_{ij} \,h_{kl}+\mathrm{perms.})\,,
\end{align}
and so on. At a first sight this is hardly going to help, since we are still dealing with an infinite series of terms that diverge when $B\gg1$.
However, one can check that $h_{ij}$ has degree $i+j-2$, $h_{ijk}$ has degree $i+j+k-4$, and similarly for higher order ones, so that $W(B;s)$ is a better behaved expansion when $B$ is large. Moreover, thanks to the exponential representation, truncating the expansion at any order is guaranteed to return a positive definite probability distribution.

This result has a nice interpretation in terms of Feynman diagrams. If one assigns a power of $B$ to each external leg and uses $(-1)^n\!\avg{\!\Delta^n\!}_c/n!$ as vertices and -1 as propagator, each Hermite polynomial in Eq.~\eqref{eq:pdf} represents the sum of all possible ways to connect the vertices listed in its coefficient, with all possible combinations of external and internal lines and the correct combinatorial factors. For instance, $\avg{\!\Delta^3\!}_cH_3(B)$ represents the one tree-level graph with three external legs (whence $B^3$) and the three one-loop graphs with one external leg (whence $-3B$) containing just one cubic vertex. In this language, $W$ becomes the generator of the connected graphs; these are obtained removing from each $H_{ijk\cdots}$ all the disconnected pieces, that is the products of two or more lower order connected terms.

In the large-$B$ limit, it is consistent to approximate this expansion keeping the leading term of each polynomial (which is equivalent to neglecting loop diagrams order by order). However, the smaller $s$ gets, the higher is the order at which one can safely truncate the expansion. Up to 4th order one recovers
\begin{equation}
  W(B;s) \simeq -\frac{B^2}{2} + \frac{\avg{\!\Delta^3\!}_c}{3!}B^3
  +\frac{\avg{\!\Delta^4\!}_c-3\avg{\!\Delta^3\!}_c^2}{4!}B^4
\end{equation}
which is enough to describe the mass function over the range of scales of interest, as discussed by \citet{damico11}.

Here, if the combinations of connected moments which appear in the expansion above were functions of $B$ only, then the resulting pdf would be self-similar in the sense used in the previous sections.  That fact that they are not, in general, functions of the scaling variable $B$,  means that the first term in equation~(\ref{eq:fNGeval}) will result in an additional contribution to $f(s)$, which must be added to equation~(\ref{sfs}).

\label{lastpage}

%%%%%%%%%%%%%%%%%%%%%%%%%%%%%%%%%%%%%%%%%%%%%%%%%%%%%%%%%%%%%%%%%%
%%%%%%%%%%%%%%%%%%%%%%%%%%%%%%%%%%%%%%%%%%%%%%%%%%%%%%%%%%%%%%%%%%

\bibliography{mybib}{}

\begin{thebibliography}{}

\bibitem[\protect\citeauthoryear{Ade et~al.,}{Ade  et~al.}{2013}]{planckNG}
Planck Collaboration, Ade P.,  et~al., 2013, \eprint{1303.5084}

\bibitem[\protect\citeauthoryear{Bernardeau, Colombi, Gazta\~naga \&
  Scoccimarro}{Bernardeau et~al.}{2002}]{bcgs02}
Bernardeau F.,  Colombi S.,  Gazta\~naga E.,    Scoccimarro R.,  2002, Phys
  Rep, 367, 1, \eprint{0112551}

\bibitem[\protect\citeauthoryear{Bond, Cole, Efstathiou \& Kaiser}{Bond
  et~al.}{1991}]{bcek91}
Bond J.,  Cole S.,  Efstathiou G.,    Kaiser N.,  1991, Astrophys.J., 379, 440

\bibitem[\protect\citeauthoryear{Bond \& Myers}{Bond \& Myers}{1996}]{bm96}
Bond J.,  Myers S.,  1996, Astrophys.J.Suppl., 103, 1

\bibitem[\protect\citeauthoryear{Castorina \& Sheth}{Castorina \&
  Sheth}{2013}]{cs13}
Castorina E.,  Sheth R.~K.,  2013, MNRAS, 433, L102, \eprint{1301.5128}

\bibitem[\protect\citeauthoryear{Corasaniti \& Achitouv}{Corasaniti \&
  Achitouv}{2011}]{ca11}
Corasaniti P.,  Achitouv I.,  2011, Phys.Rev.Lett., 106, 241302,
  \eprint{1012.3468}

\bibitem[\protect\citeauthoryear{D'Aloisio, Zhang, Shapiro \& Mao}{D'Aloisio
  et~al.}{2013}]{D'Aloisio2013}
D'Aloisio A.,  Zhang J.,  Shapiro P.~R.,    Mao Y.,  2013, \eprint{1304.6411}

\bibitem[\protect\citeauthoryear{D'Amico, Musso, Norena \& Paranjape}{D'Amico
  et~al.}{2011}]{damico11}
D'Amico G.,  Musso M.,  Norena J.,    Paranjape A.,  2011, JCAP, 1102, 001,
  \eprint{1005.1203}

\bibitem[\protect\citeauthoryear{Despali, Tormen \& Sheth}{Despali
  et~al.}{2013}]{dts13}
Despali G.,  Tormen G.,    Sheth R.~K.,  2013, MNRAS, 431, 1,
  \eprint{1212.4157}

\bibitem[\protect\citeauthoryear{Joudaki, Dor\'e, Ferramacho, Kaplinghat \&
  Santos}{Joudaki et~al.}{2011}]{Joudaki11}
Joudaki S.,  Dor\'e O.,  Ferramacho L.,  Kaplinghat M.,    Santos M.~G.,  2011,
  Phys.Rev.Lett., 107, 131304, \eprint{1105.1773}

\bibitem[\protect\citeauthoryear{Lam \& Sheth}{Lam \& Sheth}{2008}]{ls08}
Lam T.~Y.,  Sheth R.~K.,  2008, MNRAS, 386, 407, \eprint{0711.5029}

\bibitem[\protect\citeauthoryear{Lam \& Sheth}{Lam \& Sheth}{2009}]{ls09}
Lam T.~Y.,  Sheth R.~K.,  2009, MNRAS, 398, 2143, \eprint{0905.1702}

\bibitem[\protect\citeauthoryear{Lidz, Baxter, Adshead \& Dodelson}{Lidz
  et~al.}{2013}]{lidz13}
Lidz A.,  Baxter E.~J.,  Adshead P.,    Dodelson S.,  2013, Phys.Rev., D88,
  023534, \eprint{1304.8049}

\bibitem[\protect\citeauthoryear{LoVerde, Miller, Shandera \& Verde}{LoVerde
  et~al.}{2008}]{lmsv08}
LoVerde M.,  Miller A.,  Shandera S.,    Verde L.,  2008, JCAP, 0804, 014,
  \eprint{0711.4126}

\bibitem[\protect\citeauthoryear{LoVerde \& Smith}{LoVerde \&
  Smith}{2011}]{LoVerde11}
LoVerde M.,  Smith K.~M.,  2011, JCAP, 1108, 003, \eprint{1102.1439}

\bibitem[\protect\citeauthoryear{Maggiore \& Riotto}{Maggiore \&
  Riotto}{2010}]{mr10c}
Maggiore M.,  Riotto A.,  2010, Astrophys.J., 717, 526, \eprint{0903.1251}

\bibitem[\protect\citeauthoryear{Matarrese, Verde \& Jimenez}{Matarrese
  et~al.}{2000}]{MVJ00}
Matarrese S.,  Verde L.,    Jimenez R.,  2000, Astrophys.J., 541, 10,
  \eprint{astro-ph/0001366}

\bibitem[\protect\citeauthoryear{Musso \& Paranjape}{Musso \&
  Paranjape}{2012}]{mp11}
Musso M.,  Paranjape A.,  2012, MNRAS, 420, 369, \eprint{1108.0565}

\bibitem[\protect\citeauthoryear{Musso \& Sheth}{Musso \& Sheth}{2012}]{ms12}
Musso M.,  Sheth R.~K.,  2012, MNRAS, 423, L102, \eprint{1201.3876}

\bibitem[\protect\citeauthoryear{Musso \& Sheth}{Musso \& Sheth}{2013}]{ms13b}
Musso M.,  Sheth R.~K.,  2013, MNRAS, \eprint{1306.0551}

\bibitem[\protect\citeauthoryear{Musso \& Sheth}{Musso \& Sheth}{2014}]{ms14}
Musso M.,  Sheth R.~K.,  2014, to appear

\bibitem[\protect\citeauthoryear{Paranjape, Lam \& Sheth}{Paranjape
  et~al.}{2012}]{pls12}
Paranjape A.,  Lam T.~Y.,    Sheth R.~K.,  2012, MNRAS, 420, 1429,
  \eprint{1105.1990}

\bibitem[\protect\citeauthoryear{Paranjape \& Sheth}{Paranjape \&
  Sheth}{2012}]{ps12}
Paranjape A.,  Sheth R.~K.,  2012, MNRAS, 426, 2789, \eprint{1206.3506}

\bibitem[\protect\citeauthoryear{{Paranjape}, {Sheth} \&
  {Desjacques}}{{Paranjape} et~al.}{2013}]{psd13}
{Paranjape} A.,  {Sheth} R.~K.,    {Desjacques} V.,  2013, MNRAS, 431, 1503,
  \eprint{1210.1483},
  \adsurl{http://adsabs.harvard.edu/abs/2013MNRAS.431.1503P}

\bibitem[\protect\citeauthoryear{Press \& Schechter}{Press \&
  Schechter}{1974}]{ps74}
Press W.~H.,  Schechter P.,  1974, ApJ, 187, 425

\bibitem[\protect\citeauthoryear{Sheth, Chan \& Scoccimarro}{Sheth
  et~al.}{2013}]{scs13}
Sheth R.~K.,  Chan K.~C.,    Scoccimarro R.,  2013, Phys.Rev., D87, 083002,
  \eprint{1207.7117}

\bibitem[\protect\citeauthoryear{Sheth, Mo \& Tormen}{Sheth
  et~al.}{2001}]{smt01}
Sheth R.~K.,  Mo H.,    Tormen G.,  2001, MNRAS, 323, 1,
  \eprint{astro-ph/9907024}

\bibitem[\protect\citeauthoryear{Sheth \& Tormen}{Sheth \& Tormen}{2002}]{st02}
Sheth R.~K.,  Tormen G.,  2002, Mon.Not.Roy.Astron.Soc., 329, 61,
  \eprint{astro-ph/0105113}

\end{thebibliography}

\end{document}